\documentclass[fleqn,usenatbib, onecolumn]{mnras}

\usepackage{newtxtext,newtxmath}

\usepackage[T1]{fontenc}
\usepackage{ae,aecompl}

\usepackage{graphicx}	
\usepackage{amsmath}	
\usepackage{pdflscape}
\usepackage{subcaption}
\usepackage{caption}
\usepackage[utf8]{inputenc}
\usepackage{simpler-wick}
\graphicspath{{images/}}

\newcommand{\HI}{\ion{H}{i}}
\newcommand{\hi}{\ion{H}{i}~}

\title[Delensing with the CIB: impact of foregrounds]{Delensing the CMB with the cosmic infrared background: the impact of foregrounds}

\author[A. Baleato Lizancos et al.]{
    Antón Baleato Lizancos,$^{1,2}$\thanks{E-mail: a.baleatolizancos@berkeley.edu}
    Anthony Challinor,$^{1,2,3}$
    Blake D. Sherwin,$^{2,3}$ \newauthor 
    Toshiya Namikawa$^{3}$
\\
$^{1}$Institute of Astronomy, Madingley Road, Cambridge, CB3 0HA, UK\\
$^{2}$Kavli Institute for Cosmology Cambridge, Madingley Road, Cambridge, CB3 0HA, UK\\
$^{3}$DAMTP, Centre for Mathematical Sciences, University of Cambridge, Wilberforce Road, Cambridge, CB3 0WA, UK
}

\date{Accepted XXX. Received YYY; in original form ZZZ}

\pubyear{2015}

\begin{document}
\label{firstpage}
\pagerange{\pageref{firstpage}--\pageref{lastpage}}
\maketitle

\begin{abstract}
The most promising avenue for detecting primordial gravitational waves from cosmic inflation is through measurements of degree-scale CMB $B$-mode polarisation. This approach must face the challenge posed by gravitational lensing of the CMB, which obscures the signal of interest. Fortunately, the lensing effects can be partially removed by combining high-resolution $E$-mode measurements with an estimate of the projected matter distribution. For near-future experiments, the best estimate of the latter will arise from co-adding internal reconstructions (derived from the CMB itself) with external tracers such as the cosmic infrared background (CIB). In this work, we characterise how foregrounds impact the delensing procedure when CIB intensity, $I$, is used as the matter tracer. We find that higher-point functions of the CIB and Galactic dust such as $\langle BEI \rangle_{c}$ and $\langle EIEI \rangle_{c}$ can, in principle, bias the power spectrum of delensed $B$-modes. To quantify these, we first estimate the dust residuals in currently-available CIB maps and upcoming, foreground-cleaned Simons Observatory CMB data. Then, using non-Gaussian simulations of Galactic dust -- extrapolated to the relevant frequencies, assuming the spectral index of polarised dust emission to be fixed at the value determined by Planck -- we show that the bias to any primordial signal is small compared to statistical errors for ground-based experiments, but might be significant for space-based experiments probing very large angular scales. However, mitigation techniques based on multi-frequency cleaning appear to be very effective. We also show, by means of an analytic model, that the bias arising from the higher-point functions of the CIB itself ought to be negligible.
\end{abstract}



\section{Introduction}
The cosmological model best fitting observations features a period of accelerated expansion at very early times known as \emph{cosmic inflation}. In general, the physics of inflation predicts the generation during this period of a stochastic background of gravitational waves (tensor perturbations) and fluctuations in the density (scalar perturbations). Although difficult to detect directly, primordial gravitational waves are expected to have left an imprint on the temperature and polarisation of the cosmic microwave background (CMB; \citealt{Polnarev:1985}).

Scalar perturbations are now known to be responsible for the vast majority of the anisotropy pattern found in the temperature ($T$) and gradient-like $E$-mode component of polarisation. The variance associated with the scalar modes makes it impossible to detect the much smaller contribution from tensor perturbations in temperature or $E$-mode polarisation data. However, it is possible to form a curl-like component ($B$) which, at recombination (and in linear theory), is sourced only by tensor perturbations~\citep{Seljak:1996gy, Kamionkowski:1996zd}. A detection of a primordial $B$-mode is therefore widely regarded as a conclusive test of inflation, with the amplitude of the primordial power spectrum being directly related to the energy scale at which inflation occurs. This amplitude is usually parametrised by the tensor-to-scalar ratio, $r$. Currently, the most stringent experimental upper bounds are those of~\citet{ref:bicep2_18}, which place $ r<0.06 $, or $r<0.044$ in combination with Planck~\citep{ref:tristram_20}, both with 95\,\% confidence.

Efforts to constrain $r$ below this level face a significant challenge: gravitational lensing of the CMB by the large-scale distribution of matter in the Universe converts part of the primordial $E$-mode into $B$-modes~\citep{Zaldarriaga:1998ar}. The angular power spectrum of the lensing $B$-modes resembles that of white noise with $\Delta_{P}=5\, \mu \text{K\,arcmin}$ on the large angular scales where the primordial signal is expected to be strongest (an effect first detected by~\citealt{ref:hanson_13}). Consequently, the variance due to lensing becomes a significant hurdle for detecting small primordial signals. In fact, the lensing noise is at a level where it is already comparable to current experimental noise levels, and thus upcoming surveys, in addition to improving experimental sensitivity, ought to focus on removing the lensing noise.

The removal of lensing $B$-modes cannot be achieved via multi-frequency cleaning, for lensing is achromatic. However, the lensing $B$-modes can be estimated by combining observations of $E$-mode polarisation with estimates of the lensing deflections on the sky in a way that mimics how the lensing operation occurs in nature. This can then be used to remove the lensing effects in a procedure known as \emph{delensing}. This technique has already been demonstrated on real data by several teams~\citep{ref:carron_17, ref:manzotti_delensing, ref:planck_2018_lensing, ref:polarbear_delensing_19, ref:han_20, ref:bicep_delensing}.

In order to undo the deflections induced by lensing, an estimate of the projected matter distribution on the sky -- which determines the lensing convergence, $\kappa$ -- is required. For the sensitivities and resolution of current and near-future CMB experiments, the best possible estimate is obtained from tracers external to the CMB such as galaxy surveys or the cosmic infrared background~\citep{ref:smith_12,ref:sherwin_2015}, as these provide access to small-scale lenses at high redshift that internal reconstruction from the CMB itself (via the quadratic estimators of \citealt{ref:okamoto_hu_2003} or maximum-likelihood approaches in the style of \citealt{ref:hirata_03_polarization}) cannot yet extract with sufficient precision.

In recent years, one tracer has emerged as being particularly useful for delensing -- the cosmic infrared background (CIB). The CIB is the integrated emission from dust heated by UV starlight in faraway, star-forming galaxies. It originates from a broad redshift range centred around $z\approx 1$--$2$, though extending as far back as $z \approx 4$~\citep{ref:hermes_bethermin, ref:hermes_viero_wang, ref:hermes_viero_moncelsi}. This distribution overlaps extensively with the lenses responsible for CMB lensing;
so much so that the CIB and CMB lensing convergence are up to $80\,\%$ correlated \citep{ref:song_2003,ref:holder_13, ref:planck_lensing_cib_cross}. The constituent dust is at a temperature of a few tens of Kelvin so CIB emission peaks in the sub-millimetre part of the electromagnetic spectrum, where it was first detected by~\citet{ref:puget_96}, and dominates over other components (at least at high Galactic latitudes) in the high-frequency channels of CMB experiments, with lower frequencies probing higher-redshift sources.

Recent observations at 353, 545 and 857\,GHz made by the High-Frequency Instrument (HFI) aboard the \emph{Planck} satellite have led to significant improvements in our understanding of the CIB thanks to the instrument's high sensitivity and fine angular resolution and large sky coverage. On the one hand, these have allowed for a refinement of models \citep{ref:planck_2011_cib_ps,ref:planck_13_cib,ref:serra_14,ref:mak_17,ref:maniyar_18}, and on the other, high-fidelity maps of the CIB anisotropies over large fractions of the sky have been produced \citep{ref:gnilc, ref:lenz_19}. These maps are some of the best tracers of the large-scale structure of the Universe at high redshift that have been produced to date. As such, they hold great promise for upcoming implementations of delensing, particularly in combination with other tracers \citep{ref:yu_17}. CIB maps from Planck were used by \citet{ref:larsen_16} in the first demonstration of delensing, and more recently \citet{ref:bicep_delensing} used them to show the first improvement in $r$ constraints thanks to delensing. Meanwhile, \citet{ref:manzotti_delensing} used \emph{Herschel} CIB data to achieve the first statistically-significant delensing of $B$-modes.

The aim of delensing is to reduce the variance on estimates of the $B$-mode power spectrum by subtracting the specific realisation of lensing $B$-modes present on the sky. The more effective this procedure, the more attention one must pay to potential systematic errors. In spite of this, the impact of foreground emission on delensing has remained largely unexplored, with the exception of \citet{ref:beck_2020}'s recent study of the impact of residual foregrounds on internal delensing of $B$-modes. In this paper, we focus on the impact of foregrounds when using the CIB to delens $B$-modes. The principal worry is that, as part of the delensing procedure, residual amounts of dust or CIB emission left over in the CMB maps can couple with residual dust or the CIB in the matter tracer maps. The problem is exacerbated by the fact that both the CIB and Galactic dust emission are the product of radiating dust and therefore display very similar spectral energy distributions, making it difficult to disentangle the two. We show that these couplings are indeed present, even after foreground cleaning, and could be significant on the very largest angular scales. We also discuss mitigation techniques, which fortunately prove very effective.

The purpose of this paper is to gauge how delensing performance is impacted by these considerations. We begin, in section~\ref{sec:motivation}, with a brief introduction to $B$-mode delensing and how foregrounds can impact the procedure. In section~\ref{sec:methods}, the simulations of Galactic dust and CIB used in this work are described (with further details on the Gaussian CIB simulations provided in appendix~\ref{appendix:simulating_correlated_tracers}), the levels of dust residuals expected from upcoming experiments are estimated, the masking schemes we use are motivated and the overall analysis is laid out in detail. A fast implementation of a curved-sky, linear-order $B$-mode template is explained in appendix~\ref{sec:curved_sky_template}. Then, in section~\ref{sec:results}, we present our results for the magnitude of delensing bias that we expect to arise from higher-point functions of both residual Galactic dust and the CIB itself, with details of the latter provided in appendices~\ref{appendix:BET_bispectrum},~\ref{sec:equal_z_bispectrum},~\ref{appendix:toy_model_cib_higherpointfns},~\ref{appendix:discrete_bispectrum} and~\ref{appendix:ETET_trispectrum}. These results are validated in section~\ref{sec:validation} by comparing to data from Planck and to Gaussian simulations. Finally, we summarise our conclusions in section~\ref{sec:conclusion}. Throughout this paper, we assume a cosmology with parameters as determined by~\citet{ref:planck_18_legacy}.

\section{Motivation}\label{sec:motivation}
\subsection{B-mode delensing}

The lensing-induced $B$-modes can be approximated, to leading order in the lensing convergence $\kappa$ and in the flat-sky approximation, by
\begin{equation}
\label{eqn:lensB}
B^{\text{lens}}(\bmath{l}) = \int \frac{d^2 \bmath{l}'}{(2\pi)^2} W(\bmath{l},\bmath{l}') E(\bmath{l}') \kappa(\bmath{l}-\bmath{l}') \, ,
\end{equation}
where $E(\bmath{l})$ is the unlensed $E$-mode polarisation. The geometric coupling
\begin{equation}
W(\bmath{l},\bmath{l}') = \frac{2 \bmath{l}' \cdot (\bmath{l}-\bmath{l}')}{|\bmath{l}-\bmath{l}'|^2} \sin 2(\psi_{\bmath{l}} - \psi_{\bmath{l}'}) \, ,   
\end{equation}
where $\psi_{\bmath{l}}$ is the angle that $\bmath{l}$ makes with the $x$-axis. It was shown by \citet{ref:challinor_05} that the power spectrum of this linear-order approximation is in excellent agreement with the non-perturbative power spectrum
at multipoles $l\leq 2000$, which comfortably includes the range we are interested in ($l<200$) when searching for $B$-modes from primordial gravitational waves.

Given some tracer $I$ of the lensing convergence, we can construct an estimate of the lensing $B$-modes as follows:
    \begin{equation}\label{eqn:template}
    \hat{B}^{\mathrm{lens}}(\bmath{l}) = \int \frac{d^2\bmath{l}'}{(2\pi)^2} f(\bmath{l}, \bmath{l}') W(\bmath{l},\bmath{l}') E^{\mathrm{obs}}(\bmath{l}')I(\bmath{l}-\bmath{l}') \, ,
    \end{equation}
    where $E^{\mathrm{obs}}$ is the observed (lensed and noisy but beam-deconvolved) $E$-mode and $f(\bmath{l}, \bmath{l}')$ is a function that can be chosen to minimise the variance of the power spectrum of the delensed $B$-modes. We shall often refer to equation~\eqref{eqn:template} as the ``lensing template''. The delensed $B$-modes are obtained by subtracting this lensing template from the observed $B$-modes. Delensing methods structured around the construction and subtraction of a lensing template provide significant advantages over alternative approaches relying on an inverse deflection of the observed fields. This is particularly true for ground-based experiments, for which it is much more cost-effective to carry out the large-angular-scale $B$-mode science using arrays of small-aperture telescopes (SATs)\footnote{The main challenge for making highly sensitive measurements of CMB polarisation on the largest angular scales ($l<20$) is control of temperature-to-polarisation leakage of the spatially- and temporally-varying emission from the atmosphere. A leading way to achieve this is by means of devices that modulate the observed polarisation. At the time of writing, it is more cost-effective to increase sensitivity by installing these devices on many SATs rather than on a LAT~\citep{ref:s4}.}, while the high-resolution polarisation observations required to delens are made by a large-aperture telescope (LAT). The template formalism provides the necessary framework to combine the data from the different telescopes. In the limit of noiseless $E$-modes and perfect $\phi$, such a template built from lensed $E$-modes (which turns out to be preferable over using delensed $E$-modes, or forming a non-perturbative template by remapping lensed $E$-modes) can be used to reduce the residual lensing power down to around $1\,\%$ of its original amplitude. In practical applications, the method will be effectively optimal even beyond the era of CMB-S4~\citep{ref:limitations_paper}.
    
    The matter tracer, $I$, can be obtained either internally from the observed CMB or externally via observables known to correlate with the distribution of matter in the Universe. Although, ultimately, internal techniques based on quadratic estimators \citep{ref:okamoto_hu_2003}, maximum-likelihood~\citep{ref:hirata_03_polarization} or Bayesian/iterative approaches~\citep{ref:carron_lewis_map_17,ref:millea_19} will provide the best means of extracting the information, current and near-future experiments will benefit greatly from combining internal reconstructions with external tracers~\citep{ref:smith_12,ref:sherwin_2015,ref:yu_17,ref:manzotti_17,ref:bicep_delensing}. The reason for this is that large-scale lensing $B$-modes are produced from lenses over a wide range of multipoles (with 50\,\% of the power coming from lenses at $l>400$;~\citealt{ref:smith_12, ref:CORElensing}). Internal reconstructions on intermediate and small scales from experiments surveying wide fields at high angular resolution, such as the forthcoming Simons Observatory (SO;~\citealt{ref:SO_science_paper}), will be noise-dominated on these intermediate and small scales and so will benefit from combining with external tracers. Looking further ahead, for experiments such as CMB-S4~\citep{ref:s4}, the gain in delensing efficiency from adding external tracers is marginal. However, it will be possible to use the CIB in combination with other tracers as a cross-check on internal delensing.
    
    For an external tracer, \citet{ref:sherwin_2015} showed that the choice of $f(\bmath{l}, \bmath{l}')$ in equation~\eqref{eqn:template} that minimises the power spectrum of the delensed $B$-modes is a product of Wiener filters:
    \begin{equation}
        f(\bmath{l}, \bmath{l}') =  \mathcal{W}^{E}_{l'} \mathcal{W}^{I}_{|\bmath{l} - \bmath{l}'|}\, ,
    \end{equation}
     with
    \begin{equation}\label{eqn:wiener_filters}
    \mathcal{W}^{E}_l\equiv \frac{\tilde{C}_l^{EE}}{\tilde{C}_l^{EE}+N_l^{EE}} \quad \text{and} \quad
    \mathcal{W}^{I}_l \equiv \frac{C_l^{\kappa I, \mathrm{fid}}}{C_l^{II}} \, .
    \end{equation}
    Here, $\tilde{C}_l^{EE}$ is the lensed $E$-mode power spectrum, $N_l^{EE}$ is the instrument noise power, $C_l^{\kappa I}$ is the cross-power spectrum between the tracer $I$ and the true CMB lensing convergence $\kappa$, and $C_l^{II}$ is the total power spectrum of the tracer including, for example, shot noise. In the remainder of this work, we shall assume that $E$-mode observations are sample-variance limited up to $l\approx2000$, so that $\mathcal{W}^{E}_{l<2000}=1$. This is a good approximation for current and upcoming experiments with beams of a few arcmin and polarisation sensitivities better than $10\,\mu$K\,arcmin.
    
    It is worth noting that delensing performance can be further improved by linearly combining many lensing tracers $\{I_i\}$ in a way that maximises the cross-correlation between the emerging, co-added tracer and the true convergence. \citet{ref:sherwin_2015} show that the weights that achieve this are the Wiener filter such that
    \begin{equation}\label{eqn:multitracer_weights}
    \mathcal{W}_l^I I(\bmath{l}) \rightarrow \sum_{ij} I_i(\bmath{l}) \left(C_l^{-1}\right)^{ij} C_l^{\kappa I_j} \, ,   
    \end{equation}
    where $C_l^{\kappa I_i}$ is the cross-power spectrum between the true convergence and the $i$th tracer, and $\left(C_l^{-1}\right)^{ij}$ is the matrix inverse of the cross-power spectra $C_l^{I_i I_j}$ between the tracers. The correlation coefficient of the linearly-combined tracers with the true convergence can be expressed in terms of the individual correlation coefficients, $\rho_l^{\kappa I_i}$, and the correlation coefficients between tracers, $\rho_l^{I_i I_j}$, as
    \begin{equation}
    \rho_l^2 = \sum_{ij} \left(\rho_l^{-1}\right)^{ij} \rho_l^{\kappa I_i} \rho_l^{\kappa I_j} \, . 
    \end{equation}
    Qualitatively, on a given angular scale, this scheme gives most weight to those tracers that best correlate with the underlying convergence. In practice, this means that internal reconstructions, which accurately reconstruct lensing on the largest angular scales, can be supplemented with external tracers on the small scales where the former are dominated by reconstruction noise. For SO, for example, information from the CIB and photometric galaxy surveys will enable the co-added tracer to maintain a high degree of correlation with the true lensing convergence for multipoles $250\lesssim l\lesssim1000$~\citep{ref:SO_delensing_paper}, despite internal reconstructions only being signal-dominated for $L\lesssim 250$~\citep{ref:SO_science_paper}. This is of particular importance for delensing, since, as noted above, it is those intermediate and small-scale lenses that are most relevant for this purpose~\citep{ref:smith_12_external}. Since the small-scale lenses are located at high redshift (see figure~3 of ~\citealt{ref:lewis_challinor_review}), delensing will remain a challenging effort for the foreseeable future.

    Though the analysis in this paper will be restricted to the case where delensing is carried out using the CIB as the sole tracer, it will be worth keeping the multi-tracer weights in mind for two reasons. First, it will inform our choice of minimum multipole cutoff, $l_{\mathrm{min}}$, to be applied to the CIB maps -- as we will see in later sections, this choice has a significant impact on the expected amplitude of the bias in the delensed spectrum. Second, the magnitude of the bias in an analysis involving multiple tracers, one of them being the CIB, can be recovered from that determined in this work once the weight given to the CIB in the linear combination is known. In this sense, the results in this paper can be regarded as an upper limit on the amplitude of the bias terms.

    The power spectrum of residual lensing $B$-modes after delensing can be expressed to leading order as
    \begin{equation}\label{eqn:delensed_power}
        \tilde{C}_l^{BB,\mathrm{res}} = \int \frac{d^2\bmath{l}'}{(2\pi)^2} W^2(\bmath{l},\bmath{l}') C_{l'}^{EE}C_{|\bmath{l}-\bmath{l}'|}^{\kappa\kappa} \left[1- \mathcal{W}^{E}_{l'}\rho^2_{|\bmath{l}-\bmath{l}'|} \right] \, ,
    \end{equation}
    where $\rho_l$ is the correlation coefficient between the tracer at hand (in our case, the CIB, though possibly a co-added tracer) and the underlying CMB lensing convergence. Much like its lensed counterpart, the delensed spectrum above is almost constant (i.e., like white noise) on large angular scales. Furthermore, this behaviour can be shown to be robust to uncertainties in measurements of the auto- and cross-spectra (with CMB lensing) of the tracers when combined with a precise large-scale
    internal reconstruction. This means that, in actual analyses, it should be possible to deal with any uncertainties in the residual lensing power by marginalising over the amplitude of a constant spectrum. This will be further discussed in~\citet{ref:SO_delensing_paper}.However, the method is rather robust even when the CIB is the only tracer involved, with~\citet{ref:bicep_delensing} calculating that a misestimation of the cross-spectrum could bias their inference of $r$ by at most 0.2 standard deviations.

\subsection{The impact of residual foregrounds on CIB delensing}\label{sec:possible_biases}
At high frequencies, Galactic dust emission and CIB radiation both behave as grey-bodies with very similar temperatures and spectral indices. Hence, it is very difficult to separate the two based on multi-frequency information alone. Consequently, CIB maps extracted in this way -- such as those obtained by~\citet{ref:planck_14_cib_freqseparation} -- suffer considerable contamination from Galactic dust, and vice versa.

Several approaches to separating Galactic dust from the CIB have been explored in the literature. In this paper, we evaluate two such techniques. First, we consider the GNILC CIB maps of~\citet{ref:gnilc} which several groups have recently used in implementations of delensing~\citep{ref:larsen_16, ref:yu_17, ref:bicep_delensing}. This algorithm makes use of differences in the angular power spectrum of the two components -- approximately $l^{-1}$ for the CIB~\citep{ref:planck_2011_cib_ps} and $l^{-2.7}$ for the dust~\citep{ref:planck_14_cib_freqseparation} -- to separate them.

Another method to disentangle CIB and Galactic dust emission is by exploiting the correlation between 21-cm emission from neutral Hydrogen, \HI, in the inter-stellar medium (ISM) and Galactic dust~\citep{ref:planck_2011_cib_ps}. Recently, \citet{ref:lenz_19} have harnessed this to clean high-frequency Planck observations using \hi data from the \HI4PI survey~\citep{ref:H14PI}, producing high-quality CIB maps on a wide range of angular scales. As a side note, \hi gas in the ISM tends to form filamentary structures aligned with the local Galactic magnetic field~\citep{ref:clark_14, ref:clark_15}, so it correlates with polarised emission from asymmetric dust grains, which also tend to align with their short axes pointing along the local magnetic field. By harnessing this connection, observations of \hi have the potential to further our understanding of Galactic foregrounds for CMB polarisation studies.

From the considerations above, it is clear a complete removal of dust from the CIB maps is virtually impossible. Hence, we now seek to identify the ways in which the presence of residual dust could impact the power spectrum of delensed $B$-modes -- the estimator from which we hope to extract information about a primordial component. The most naïve approach would have us model this estimator as the power spectrum of residual lensing $B$-modes, calculated ignoring any dust or CIB residuals in the $B$-modes and in the $E$-modes within the lensing template, plus the power spectra of experimental noise, and residual amounts of dust and CIB $B$-mode polarisation. That is
\begin{align}\label{eq:naivepower}
    C_l^{BB\mathrm{,del,naive}} & = \tilde{C}_l^{BB,\mathrm{res}} + N_l^{BB} + C_l^{BB,\mathrm{dust, res}}+ C_l^{BB,\mathrm{CIB, res}} \nonumber \\
    & = \tilde{C}_l^{BB} - 2\, g_l\left[ \langle \tilde{B} \tilde{E}  I^{\mathrm{CIB}} \rangle \right] + h_l\left[\langle \tilde{E}   I^{\mathrm{tot}}\tilde{E}   I^{\mathrm{tot}}\rangle \right] + N_l^{BB} + C_l^{BB,\mathrm{dust, res}}+ C_l^{BB,\mathrm{CIB, res}} \,,
\end{align}
where $I^{\mathrm{tot}}$ includes CIB ($I^{\text{CIB}}$, composed of correlated signal and shot noise),
residual Galactic dust ($I^{\text{dust}}$) and experimental noise ($I^{\text{noise}}$); $\tilde{E}$ denotes the lensed $E$-modes; and $\tilde{C}^{BB}_l$, $N_l^{BB}$, $C_l^{BB,\mathrm{dust, res}}$ and $C_l^{BB,\mathrm{CIB, res}}$ are, respectively, the angular power spectra of lensing, instrument noise, residual dust and residual CIB $B$-modes. For economy of notation, we have denoted the convolution integral of equation~\eqref{eqn:template} as the functional $g_l$, written the double integral as the functional $h_l$, and hidden the multipole-dependence of their arguments\footnote{The last two terms in equation~(\ref{eq:naivepower}) are normally calculated to leading order in $\kappa$, which amounts to ignoring lensing in $\tilde{E}$ and replacing $\tilde{B}$ by the leading-order expression $\tilde{B} \sim E\phi$.
This is actually very accurate (with errors at the $1\,\%$ level of $\tilde{C}_l^{BB}$ in the case that $I^{\text{tot}} = \phi$, and presumably even better for a partially-correlated tracer). However, the individual $\phi^4$ contributions to the three- and four-point terms may be larger (at the $10\,\%$ level of $\tilde{C}_l^{BB}$), but they cancel rather precisely~\citep{ref:limitations_paper}.}. We have also assumed that $E$-mode instrument noise is negligible, but this can easily be accounted for by replacing the lensed $E$-modes in the $h_l$ term in equation~(\ref{eq:naivepower}) with their noisy counterparts. 
The inclusion of the dust and noise components along with the CIB in $I^{\text{tot}}$ in this term is in order to take into account the decorrelation between CIB and lensing convergence that comes about as a consequence of their presence. 

The four-point function in equation~(\ref{eq:naivepower}) can be written in terms of its connected part (denoted $\langle \rangle_c$) and Gaussian contractions as 
\begin{align}
\langle \tilde{E}   I^{\mathrm{tot}}\tilde{E}   I^{\mathrm{tot}}\rangle &= \langle \tilde{E}   I^{\mathrm{tot}}\tilde{E}   I^{\mathrm{tot}}\rangle_c +
\wick{\c1 {\tilde{E}} \c2 I^{\text{tot}} \c1 {\tilde{E}} \c2 I^{\text{tot}}} \nonumber \\
&= \langle \tilde{E}   I^{\mathrm{CIB}}\tilde{E}   I^{\mathrm{CIB}} \rangle_c + 
\wick{\c1 {\tilde{E}} \c2 I^{\text{tot}} \c1 {\tilde{E}} \c2 I^{\text{tot}}} \, ,
\end{align}
where we have assumed that all fields are zero-mean. In the Gaussian contractions we have used the fact that the lensed $E$-modes are uncorrelated (although not independent) of $I^{\text{tot}}$. 

In reality, there will also be residual amount of Galactic dust and CIB contained in the $E$-mode data, as well as residual dust in the CIB map. These will give rise to additional contributions not included in equation~\eqref{eq:naivepower} These possible additional terms -- which we will be referring to as ``biases'' -- are schematically described by
\begin{align}\label{eqn:bias_def}
    \mathcal{B}_l & \equiv C_l^{BB,\mathrm{del}} - C_l^{BB,\mathrm{del, naive}} \nonumber \\
    & = - 2\, g_l\left[ \langle B^{\mathrm{dust}} E^{\mathrm{dust}}  I^{\mathrm{dust}} \rangle_{c} + \langle B^{\mathrm{CIB}} E^{\mathrm{CIB}}  I^{\mathrm{CIB}} \rangle_{c} \right] + h_l\left[ \langle E^{\mathrm{dust}}  I^{\mathrm{dust}} E^{\mathrm{dust}}  I^{\mathrm{dust}}\rangle + 
    \wick{\c1 E^{\mathrm{dust}}  \c2 I^{\mathrm{CIB}} \c1 E^{\mathrm{dust}} \c2 I^{\mathrm{CIB}}}
    + \wick{\c1 E^{\mathrm{dust}}  \c2 I^{\mathrm{noise}} \c1 E^{\mathrm{dust}} \c2 I^{\mathrm{noise}}}\right] \nonumber \\
    & \mbox{} + h_l\left[\langle E^{\mathrm{CIB}}  I^{\mathrm{CIB}} E^{\mathrm{CIB}}  I^{\mathrm{CIB}}\rangle_{c} + 
    \wick{\c1 E^{\mathrm{CIB}}  \c2 I^{\mathrm{tot}} \c1 E^{\mathrm{CIB}} \c2 I^{\mathrm{tot}}} \right] \, .
\end{align}
Here, we have ignored terms involving the two-point function $\langle E^{\text{CIB}} I^{\text{CIB}}\rangle$ since this 
vanishes if we assume, as we shall do, that the polarisation angles of CIB galaxies are uncorrelated (see section~\ref{sec:analytic_cib_bias_calc}). Of the three terms with the Gaussian contractions shown explicitly, we shall only consider the first (i.e., that with $I^{\text{CIB}}$ shown explicitly), since we can ignore instrument noise in the CIB measurement on the scales relevant for delensing and since CIB polarisation residuals are small relative to the polarised dust residuals \citep{ref:lagache_19}. This work estimates the remaining terms. Notably, that includes bispectra and trispectra of Galactic dust and CIB given the non-Gaussian statistics of these emission components.

In section~\ref{sec:bias_from_sims}, we use simulations to calculate the bias resulting from $g_l[ \langle B^{\mathrm{dust}} E^{\mathrm{dust}}  I^{\mathrm{dust}} \rangle_{c}]$, $h_l[\langle E^{\mathrm{dust}}  I^{\mathrm{dust}} E^{\mathrm{dust}}  I^{\mathrm{dust}}\rangle]$ and $h_l[\langle E^{\mathrm{dust}} I^{\mathrm{CIB}} E^{\mathrm{dust}} I^{\mathrm{CIB}}\rangle]$. The dust simulations we use are detailed in section~\ref{sec:dust_sims}, while those of the CIB are explained in section~\ref{sec:cib_sims}. At the time of writing there exist no simulations of the polarised component of the CIB, so an analytic treatment of $\langle B^{\mathrm{CIB}} E^{\mathrm{CIB}}  I^{\mathrm{CIB}} \rangle_{c}$ and $\langle E^{\mathrm{CIB}}  I^{\mathrm{CIB}} E^{\mathrm{CIB}}  I^{\mathrm{CIB}}\rangle_{c}$ is required. This calculation is presented in section~\ref{sec:analytic_cib_bias_calc}.

In addition to introducing a bias, the presence of residual foregrounds will increase the variance of the delensed power spectrum. Part of this additional variance will be associated with tracer fluctuations which can be described using Gaussian statistics; so when actually doing the analysis, this effect could be easily captured using simulations. However, there will be additional contributions to the variance coming from the non-Gaussianity of the foregrounds. When estimating the impact of CIB non-Gaussianities, progress can be made with existing simulations (see, e.g.,~\citealt{ref:bicep_delensing}). In the case of polarised Galactic dust, this will also require generating non-Gaussian simulations, or obtaining a faithful template from observational data. Though important, the study of this additional variance is beyond the scope of this work: in what follows, we restrict ourselves to assessing the impact of delensing biases.

\section{Methods}\label{sec:methods}
In this section we introduce much of the machinery used in section~\ref{sec:bias_from_sims} to estimate, based on simulations, the delensing bias arising from residual Galactic dust.
\begin{table}
    \centering
    \begin{tabular}{| l | l | l |}
        \hline
        Symbol & Description & Value \\ \hline
        $T_{\mathrm{dust}}$ & Equilibrium temperature of the thermal dust SED $[\mathrm{K}]$ & 19.6 \\
        $\beta_{\mathrm{dust}}$ & Spectral index of the thermal dust emission &$1.53$ \\
        \hline
    \end{tabular}
    \caption{Dust parameters used in this work, obtained from~\citet{ref:planck_phenom_dust_model}.}\label{tab:constants}
\end{table}

\subsection{Simulations of Galactic dust}\label{sec:dust_sims}
    Given the lack of small-scale polarisation data currently available, we rely on simulations to perform our analysis. In particular, we make use of a single, full-sky, non-Gaussian simulation of the Galactic dust component at 353\,GHz produced by~\citet[hereafter VS]{ref:vansyngel}. We focus on this frequency for two reasons: first, it is the highest frequency for which there is polarisation data from Planck, which will enable the validation process described in section~\ref{sec:validation}; and second, that, among the high-frequency Planck channels, the one at 353\,GHz is where the CIB is brightest relative to Galactic dust (see, e.g., ~\citealt{ref:mak_17}), so that one might \emph{a priori} expect CIB maps at this frequency to be particularly useful for delensing.

    The phenomenological model underlying this simulation is an extension of that in~\citet{ref:planck_phenom_dust_model}, which models the local Galactic magnetic field (GMF) by stacking an ordered component together with several layers containing Gaussian random realisations of turbulence drawn from a power-law angular power spectrum. The free parameters of the model are determined by fitting the model spectra to the Planck data. VS then builds on this to allow for a level of $TE$ correlation and $E/B$ power asymmetry that matches the observations of~\citet{ref:planck_18_polarised_dust}. We emphasise that the total intensity in this simulation is based on dust inferred from Planck observations, and so is representative of (and correlated with) our sky at the map level. In section~\ref{sec:validation}, we will validate this simulation against Planck data, and also compare it to Gaussian simulations.

    Assuming that emission from thermal dust is perfectly correlated across microwave frequencies, the dust intensity in the 145\,GHz ``science channel'' (the one where the CMB is brightest relative to the foregrounds) can be obtained by extrapolating the 353\,GHz simulated template down as
    \begin{equation}\label{eqn:dust_scaling}
        I_\nu^{\mathrm{dust}}(145\,\mathrm{GHz}, \hat{\bmath{n}}) = I_\nu^{\mathrm{dust}}(353\,\mathrm{GHz}, \hat{\bmath{n}}) \left(\frac{145}{353}\right)^{\beta_{\mathrm{dust}}}\frac{B(145\,\mathrm{GHz}, T_{\mathrm{dust}})}{B(353\,\mathrm{GHz}, T_{\mathrm{dust}})}\,,
    \end{equation}{}
    and similarly for $Q$ and $U$. This scaling assumes that the dust spectral energy distribution (SED) is well modelled as a modified blackbody~\citep{ref:thorne_17} with temperature $T_{\mathrm{dust}}$ and spectral index $\beta_{\mathrm{dust}}$ as determined by~\citet{ref:planck_18_polarised_dust}. The values of $T_{\mathrm{dust}}$ and $\beta_{\mathrm{dust}}$ used are detailed in Table~\ref{tab:constants}. The Planck function is $B(\nu,T)$. To convert this specific intensity to the corresponding differential CMB temperature at 145\,GHz, we apply a multiplicative unit conversion factor appropriate for infinitesimally-narrow frequency bands; this can be calculated using, for example, equation~(8) of~\citet{ref:planck_HFI_spectral_response}.

    The extrapolation described above ignores spatial variations in the dust temperature and spectral index that are known to be present in the data~\citep{ref:planck_18_polarised_dust}. We deem this to be an acceptable assumption for our purposes, given that our treatment already relies on a number of other approximations and imperfect simulations. However, this issue should be investigated in future work (it is, for instance, an important issue for component separation efforts). In addition to this, we also expect some further decorrelation in polarised dust emission across frequencies for any that passes through more than one dust cloud. The reason, explained by~\citet{ref:tassis}, is that, in that direction, the total $Q$ and $U$ Stokes parameters arise as a sum of contributions with different polarisation orientations (to the extent that the GMF direction threading the clouds changes along the line of sight) and whose relative strengths vary with frequency if the temperature of the clouds differs (because then their emission would follow different SEDs). However, \citet{ref:poh} have shown that this effect contributes only a small amount to the current extrapolation error budget for Planck data, and indeed~\citet{ref:planck_18_polarised_dust} find no evidence of decorrelation between 150\,GHz and 353\,GHz at the high Galactic latitudes that we will be restricting our analysis to.

\subsection{Simulations of the CIB}\label{sec:cib_sims}
The subset of  bias terms described in section~\ref{sec:possible_biases} associated with CIB emission can, in principle, be computed using simulations. However, among these terms are $\langle B^{\mathrm{CIB}} E^{\mathrm{CIB}}  I^{\mathrm{CIB}} \rangle_{c}$ and $\langle E^{\mathrm{CIB}}  I^{\mathrm{CIB}} E^{\mathrm{CIB}}  I^{\mathrm{CIB}}\rangle_{c}$, and estimating these would require non-Gaussian simulations of the CIB polarisation, which, unfortunately, are not available at the time of writing. We therefore study them analytically in section~\ref{sec:analytic_cib_bias_calc}. The remaining CIB-related bias terms can be computed using Gaussian simulations. In this section, we explain how those simulations can be generated. For the sake of wider applicability, our method allows for the resulting CIB realisations to be appropriately correlated with an underlying realisation of the lensing convergence. However, we note that having the CIB correlate with $\kappa$ is not strictly necessary when calculating the subset of bias terms we are after.

Generating correlated Gaussian simulations is straightforward if all the relevant auto- and cross-spectra are known. Suppose $\kappa_{lm}$ are the spherical harmonic coefficients of a particular realisation of the lensing convergence. We seek to simulate $I_{lm}$, the coefficients of a simulation of the CIB. (We drop the superscript CIB here to avoid cluttering the notation.)
Given that the two are correlated, we can say that, in general,
     \begin{equation}\label{eqn:cib_sim}
         I_{lm} = A_l^{I\kappa}\kappa_{lm} + u_{lm},
     \end{equation}
     where $ u_{lm}$ are the harmonic coefficients of the uncorrelated part and $A_l^{I\kappa}$ is a function to be solved for. By requiring the correct auto and cross-spectra, we see that
     \begin{equation}
         A_l^{I\kappa} = \frac{C_l^{I\kappa}}{C_l^{\kappa\kappa}},
     \end{equation}
     and the power spectrum of the uncorrelated part is
     \begin{equation}
         C_l^{uu} = C_l^{II} - (A_l^{I\kappa} )^2C_l^{\kappa\kappa}.
     \end{equation}
     We calculate $C_l^{\kappa\kappa}$ for the parameters of the best-fit $\Lambda$CDM cosmology of~\citet{ref:planck_18_legacy}. On the other hand, $C_l^{II}$ and $C_l^{I\kappa}$  are obtained following~\citet{ref:yu_17}. Briefly, $C_l^{I\kappa}$ is based on fitting the single-SED CIB model of~\citet{ref:hall_CIB} to the cross-spectrum of the GNILC CIB map (at 353\,GHz) and the internal reconstruction of $\kappa$ from~\citet{ref:planck_2015_lensing}. The same CIB model, plus a shot noise and power-law Galactic dust contribution, is then used to fit the auto-spectrum of the GNILC map. We construct $C_l^{II}$ as the sum of the modelled (clustered) CIB spectrum and shot noise.\footnote{All spectra were kindly provided by Byeonghee Yu.}

    Finally, we can generate $I_{lm}$ as described in equation~\eqref{eqn:cib_sim} by drawing the harmonic coefficients $u_{lm}$ from a Gaussian probability distribution with mean zero and variance given by $C_l^{uu}$, and adding them to a filtered version of the convergence, $A_l^{I\kappa}\kappa_{lm}$.

     In appendix~\ref{appendix:simulating_correlated_tracers}, we generalise this approach to an arbitrary number of tracers and make the implementation publicly available\footnote{\texttt{https://github.com/abaleato/MultitracerSims4Delensing}}. Along with the weighting scheme of equation~\eqref{eqn:multitracer_weights} -- also implemented -- this framework can be used to simulate multi-tracer estimates of the lensing potential with which to test delensing procedures.

\subsection{Estimating dust residuals in the data}\label{sec:dust_transfer_funcs}
    In addition to simulating the raw Galactic dust and CIB emission, we must also determine what fraction of that original radiation is left in the CMB maps after foreground cleaning, as well as how much residual dust is present in the CIB maps used as the matter tracers for delensing. In general, these residuals will be a function of angular scale and of telescope type.

\begin{figure}
        \begin{center}\includegraphics[width=0.6\columnwidth]{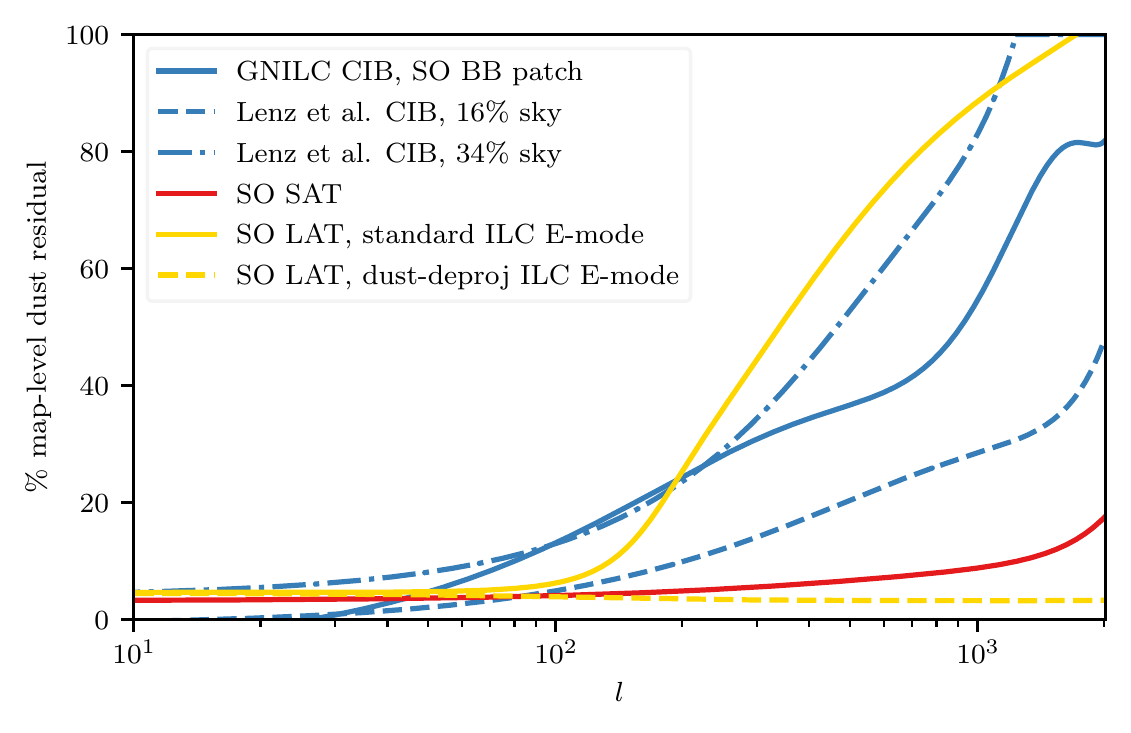}\end{center}
        \caption{Estimated map-level percentage of residual dust emission after foreground cleaning relative to the unmitigated dust emission in either the 145\,GHz channel of the SO LAT (yellow), the 145\,GHz channel of the SO SAT (red) or different CIB maps at 353\,GHz (blue). These fits to CIB residuals are not to be trusted below $l<100$; however, this is irrelevant in practical applications, since internal techniques dominate the reconstruction on those scales (see, e.g.,~\citealt{ref:SO_delensing_paper}). For the SO, we plot the \emph{goal} noise levels. The SO LAT residuals are estimated using a global ILC method and thus represent the fraction by which to scale the unmitigated emission anywhere on the sky. On the other hand, the residuals quoted for the CIB maps are appropriate to the specific regions of the sky on which they are estimated.}
        \label{fig:jpl_vs_gnilc_dust_res}
    \end{figure}

    In order to identify what residual dust fraction to expect for the CIB maps, we first investigate the GNILC maps. Their construction from the multi-frequency Planck maps is described in detail in~\citet{ref:gnilc}; see also \citet{ref:remazeilles2011}. For our purposes, the key points are as follows: first, at each frequency, an estimate of the dust component, $m_{\mathrm{dust}}$, is extracted by filtering out the CIB, CMB and instrumental noise components; then, another map, $m_{\mathrm{dust + CIB}}$, is obtained by removing only the CMB and the noise (this is done by withdrawing the prior on the CIB angular power spectrum from the procedure for obtaining $m_{\mathrm{dust}}$). The CIB maps at each frequency are then obtained by subtracting $m_{\mathrm{dust}}$ from $m_{\mathrm{dust + CIB}}$. Given that the extent of confusion between dust and noise or CMB emission ought to be relatively small, it follows that any dust that $m_{\mathrm{dust}}$ fails to capture will make it into $m_{\mathrm{dust + CIB}}$, and will therefore appear as a dust residual in the CIB map. If we assume that the multipoles of $m_{\mathrm{dust}}$ extracted over the region of interest contain a fraction $\alpha_l$ of the true dust multipoles for that region, while $m_{\mathrm{dust + CIB}}$ accurately captures the sum of the dust and CIB, the residual dust in the estimated CIB multipoles will be $1-\alpha_l$ of the raw dust level. We estimate $\alpha_l$, and hence the fractional dust residual at a given frequency,  multipole and sky region, from the cross-correlation, on the patch of choice, between GNILC maps obtained by applying the GNILC algorithm to simulations containing all sky components (specifically, the FFP8 Planck simulations described in~\citealt{ref:planck_sims}; the results were kindly provided by Mathieu Remazeilles) and the input dust in these simulations. The resulting estimate for the multipole-dependent, fractional residual at $353$\,GHz is shown in figure~\ref{fig:jpl_vs_gnilc_dust_res}. We note that the GNILC algorithm is optimised for dust reconstruction, so other methods designed to prioritise the extraction of the CIB might display higher levels of residual dust. We also point out that the GNILC algorithm tends to suppress the true CIB due to residual CIB in $m_{\mathrm{dust}}$, which is subtracted from $m_{\mathrm{dust + CIB}}$. There is evidence of this in the cross-correlation of the GNILC CIB maps with $\kappa$ from Planck, as discussed in the appendix of~\citet{ref:maniyar2019}. However, for the purposes of delensing, provided that correlation coefficients are obtained empirically any suppression of the CIB will not lead to bias but only a reduction in the delensing efficiency.

    Another means of mapping dust in our galaxy is via its correlation with the 21\,cm emission from the hyperfine transition of neutral hydrogen. This technique was first exploited by~\citet{ref:planck_2011_cib_ps} to disentangle Galactic dust from extragalactic CIB emission. Subsequently, \citet{ref:lenz_19} extended the method to larger regions of the sky, and it is those maps that we consider here. In order to estimate the magnitude of dust residuals in them, we focus on the strong correlation observed between higher \hi column densities ($N_{\text{\hi}}$) -- or, equivalently, larger sky areas -- and higher dust residuals. This is the reason why angular power spectra of maps obtained using different column density thresholds converge only for the lowest few values of $N_{\text{\hi}}$, and grow significantly in amplitude for higher ones (see figure 15 of~\citealt{ref:lenz_19}). If we assume that all that additional auto-power is due to dust contamination, we can make a rough estimate of the residual dust power present in any given CIB map by comparing its power spectrum to that of the map with the lowest column density ($N_{\text{\hi}}=1.5\times10^{20}$\,cm$^{-2}$), and assuming that the latter is free of dust. In order to avoid noise biases, auto-spectra are measured by taking the cross-correlation of maps derived from Planck observations with independent instrumental noise (the half-ring maps), masked with the appropriate apodised windows provided. As the maps from~\citet{ref:lenz_19} are provided in intensity units, the resulting absolute residual spectra are converted to temperature units by multiplying by a wide-band unit conversion factor\footnote{This multiplicative unit correction differs slightly from the one we met in section~\ref{sec:dust_sims} due to the fact that it takes into account the finite width of the Planck bandpasses -- see~\citet{ref:planck_HFI_spectral_response} for an explanation, and~\citet{ref:planck_18_polarised_dust} for values appropriate for the 2018 polarisation data we use.} twice and subsequently compared to the raw power of the VS dust simulation at $353\,$GHz. The fractional residual shown in figure~\ref{fig:jpl_vs_gnilc_dust_res} is the ratio between these two. Away from the small scales where it increases due to the limited resolution of the \hi maps, the residual dust power is approximately scale-independent. On the other hand, the raw dust power is very red, hence the shape of the residual fraction curve shown.

    We now consider the removal of dust from small-scale polarisation observations of a survey with the characteristics of the Simons Observatory's (SO) large-aperture telescope survey (LAT;~\citealt{ref:SO_science_paper}) by means of an internal linear combination of channels (ILC) in harmonic space (see, e.g.,~\citealt{ref:tegmark_03}). Though, a priori, one might think this easier than temperature cleaning, owing to the fact that in polarisation there are only two contaminants to the signal of interest (dust and synchrotron), the situation is complicated by the fact that LAT polarisation noise levels are higher than dust emission for all but the largest angular scales~\citep{ref:SO_science_paper}. Consequently, in a standard implementation of the ILC method, the most effective way to minimise the variance of the cleaned map while leaving the CMB signal unchanged will be to suppress the noise, with little being done to remove dust. Quantitatively, the fraction, $R_l$, of the original Galactic dust emission in the $145$\,GHz channel of SO that remains in the ILC-cleaned map at multipole $l$ is
    \begin{equation}
        R_{l} = \sum_{\nu} w_{l}(\nu) I^{\mathrm{dust}}(\nu)/I^{\mathrm{dust}}(145\,\mathrm{GHz}),
    \end{equation}
    where $w_{l}(\nu)$ are the ILC weights\footnote{Based on the ILC weights provided by the authors of ~\citet{ref:SO_science_paper}.}, $I^{\mathrm{dust}}(\nu)$ is the dust SED assumed in the simulations in differential CMB temperature units (such that $\sum_{\nu}w_{l}(\nu)=1$) and the sum is over all 13 of the Planck and SO frequency channels expected to be used for component separation. The unit conversions at all these frequencies are calculated under the assumption of infinitely-narrow passbands. Finally, the residual level of dust after ILC-cleaning (with noise levels corresponding to the SO's \emph{goal}) is shown in figure~\ref{fig:jpl_vs_gnilc_dust_res}.

\begin{figure}
        \begin{center}
        \includegraphics[width=0.6\columnwidth]{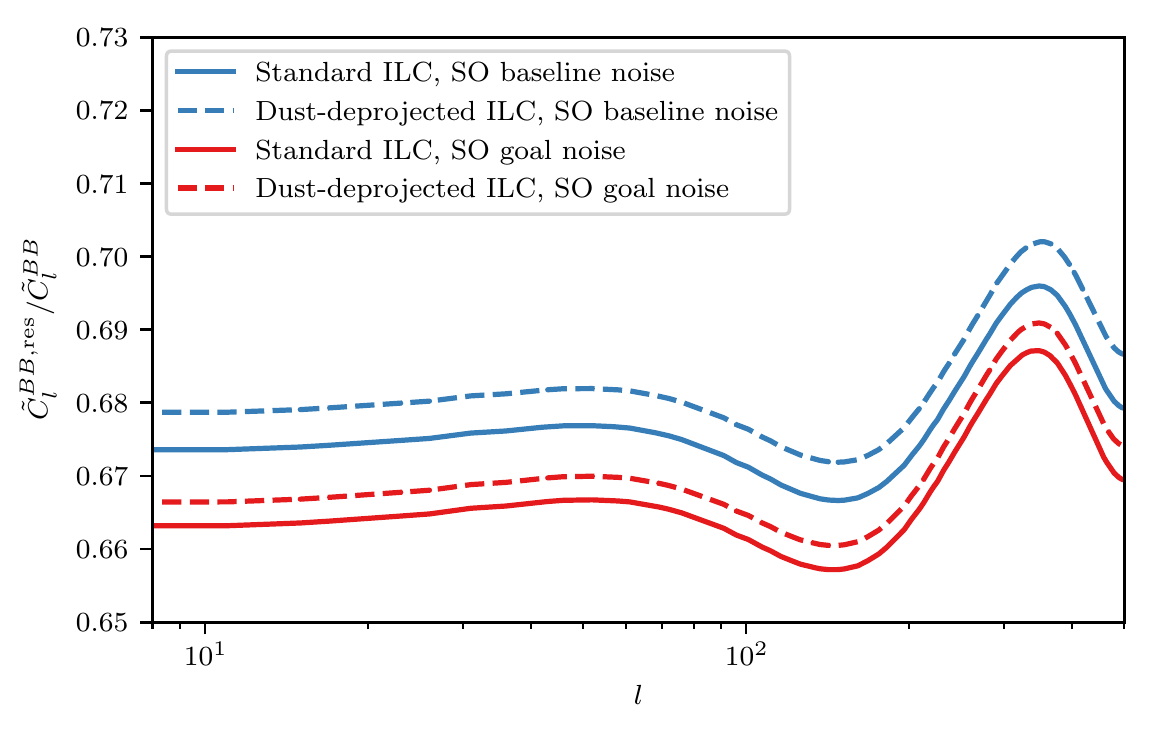}
        \end{center}
        \caption{Fraction of lensing $B$-mode power left over after delensing with the GNILC CIB map and $E$-mode observations from the SO LAT. In both the baseline and goal scenarios, nulling the dust component in the foreground-cleaned $E$-mode maps brings about a small degradation in delensing efficiency relative to the standard, minimum-variance solution. This theoretical estimate is obtained from the leading-order approximation of equation~\eqref{eqn:delensed_power}.}
        \label{fig:delensing_efficiency_after_deprojecting}
    \end{figure}

    The LAT $E$-mode foreground residuals can be further suppressed by ``deprojecting'' the dust contribution -- that is, requiring that the cleaned map have zero response to the SED of the dust component \citep{ref:remazeilles_contrained_ilc}. Though in principle one might expect this method completely to neutralise dust residuals, in practice the removal will not be complete due to differences between the dust model used to compute the ILC weights and the true dust SED (arising, for instance, from spatial variation of the dust spectral index). Our calculated dust-deprojected residual, shown in figure~\ref{fig:jpl_vs_gnilc_dust_res}, illustrates this, as it arises from differences in the dust model used to generate the ILC weights and that which we use to extrapolate the dust template from 353\,GHz to other frequencies. This estimate is likely to be rather optimistic given that it appears to be comparable (when measured as a fraction of the unmitigated emission) to that obtained in a parametric foreground cleaning of the SAT -- as we shall soon discuss, this is expected to be more extensive since it allows for a degree of variation of foreground parameters across the sky. However, our estimates will suffice given the number of other approximations we are making. As we will see in section~\ref{sec:bias_from_sims}, deprojecting the dust can help mitigate the magnitude of delensing biases. However, doing so comes at the cost of increased variance in the maps, which in turn brings about a degradation of delensing efficiency. In order to estimate the extent of the degradation, we use equation~\eqref{eqn:delensed_power} to forecast the residual lensing $B$-mode power after delensing with the GNILC CIB maps and $E$-mode observations after the latter are Wiener-filtered to reflect the polarisation noise level after ILC cleaning. Our forecasts, shown in figure~\ref{fig:delensing_efficiency_after_deprojecting}, predict a small degradation in delensing efficiency (quantified through the ratio of delensed-to-original lensing $B$-mode power), no greater than one percent, between the dust-deprojected and standard cleaning scenarios. When deciding whether or not to use dust-deprojected maps for delensing, the goal must be to maximise delensing efficiency while keeping the emerging delensing bias to acceptable levels. The newly developed partially-constrained ILC method of~\citet{ref:Abylkairov_et_al_20} might be particularly useful in this context, allowing optimisation of the tradeoff between post-cleaning noise (and associated delensing efficiency) and bias on $r$.

    Finaly, the cleaning procedure ought to be more extensive (fractionally speaking) for a small-aperture telescope -- such as the the SO's SAT -- measuring CMB polarisation on large angular scales. It is on those large scales that polarised foregrounds peak and are brighter than the experimental noise (assuming that systematics are under tight enough control that leakage of atmospheric emission into polarisation is small). Hence, they will be preferentially removed in foreground-cleaning procedures. Furthermore, given the importance of mitigating the foreground variance on large angular scales to faciliate searches for primordial $B$-modes, map-based component separation techniques have been developed that allow for variation of the foreground modelling parameters across the sky (see, e.g.,~\citealt{ref:stompor_xforecast} or~\citealt{ref:alonso_bfore} and references therein). This flexibility results in improved cleaning: by simulating the entire foreground-cleaning operation for the SAT, \citet{ref:SO_science_paper} forecast a map-level dust residual of around $10\,\%$, shown also in figure~\ref{fig:jpl_vs_gnilc_dust_res}.

    We fit smooth functions to all of these map-level dust residual fractions as a function of multipole and use them to filter the simulated, raw dust maps of section~\ref{sec:dust_sims}. The end result is maps of the expected residual dust in each of the $E,B$ and $I$ fields involved in the delensing analysis.

\subsection{Masking}
When calculating the delensing biases of section~\ref{sec:possible_biases} from simulations, we must adopt masks that mimic the procedures that would be followed and sky areas that would be covered in actual analyses.

\begin{figure}
    \begin{subfigure}[t]{0.45\textwidth}
        \centering
        \includegraphics[width=\linewidth]{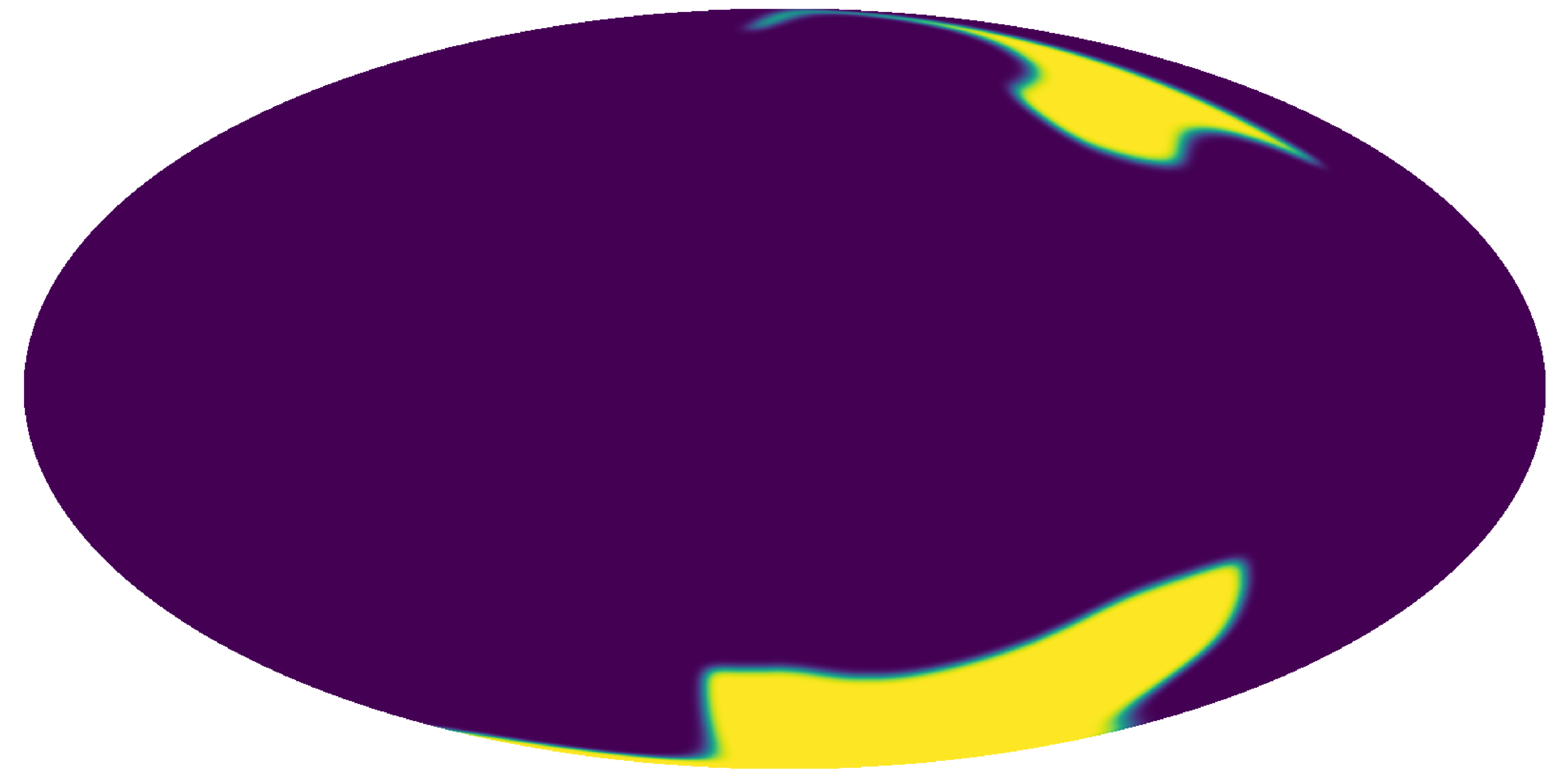}
        \caption{SO SAT mask} \label{fig:so_bb_mask}
    \end{subfigure}
    \begin{subfigure}[t]{0.45\textwidth}
        \centering
        \includegraphics[width=\linewidth]{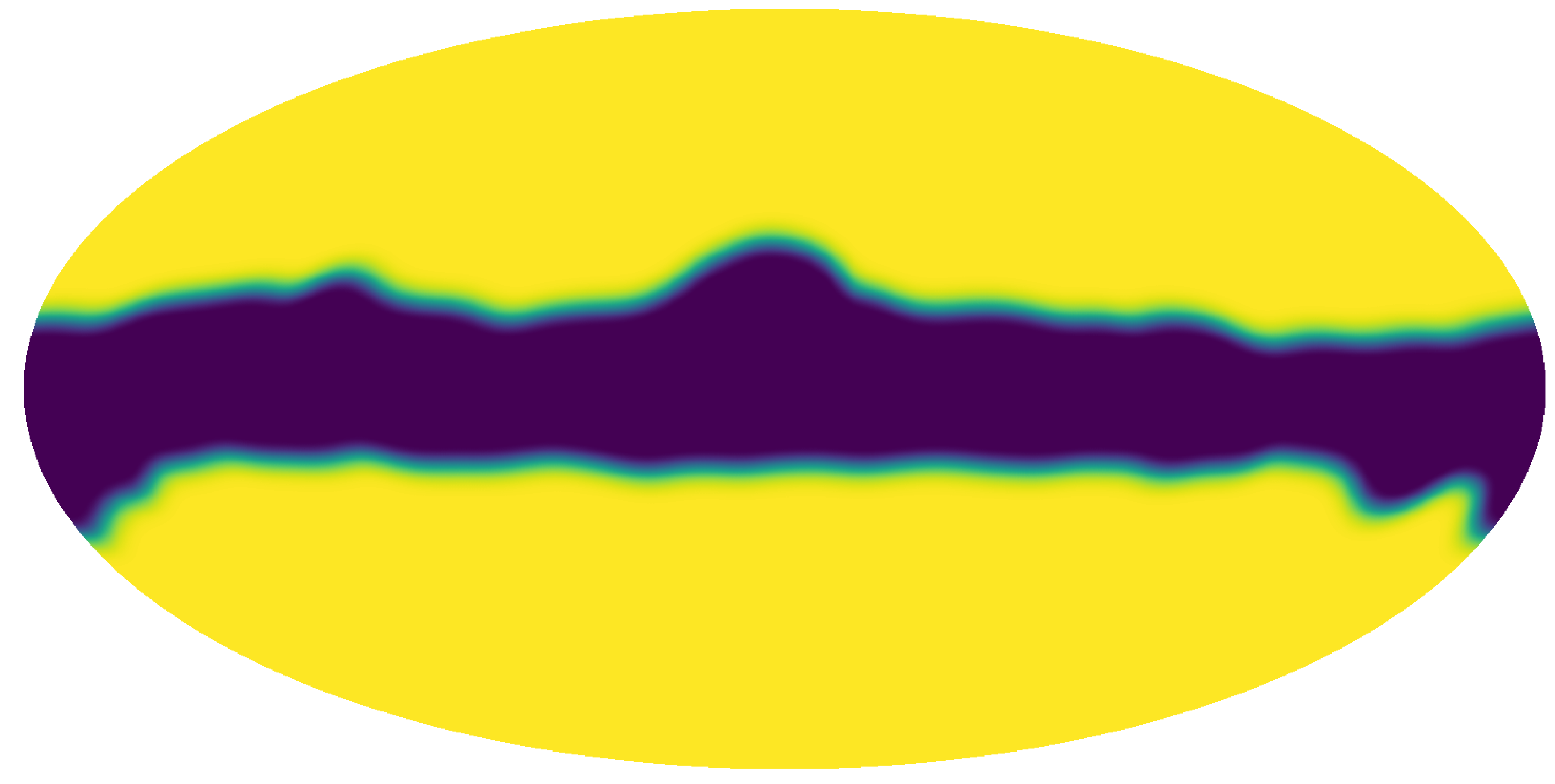}
        \caption{Planck's 80\,\% sky Galactic mask} \label{fig:planck_80pcsky_mask}
    \end{subfigure}
    \centering
    \begin{subfigure}[t]{0.45\textwidth}
        \centering
        \includegraphics[width=0.5\textwidth]{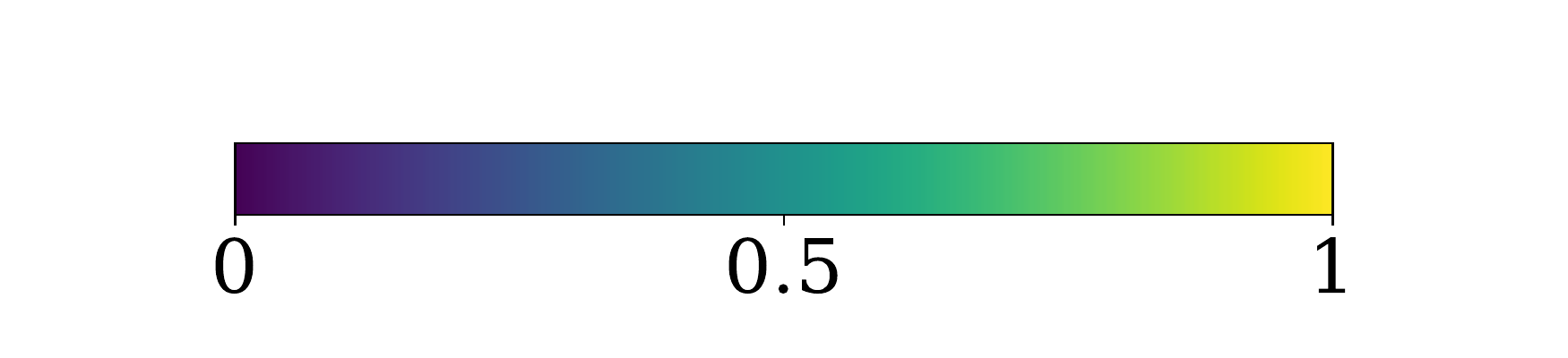}
    \end{subfigure}
    \begin{subfigure}[t]{0.45\textwidth}
        \centering
        \includegraphics[width=\linewidth]{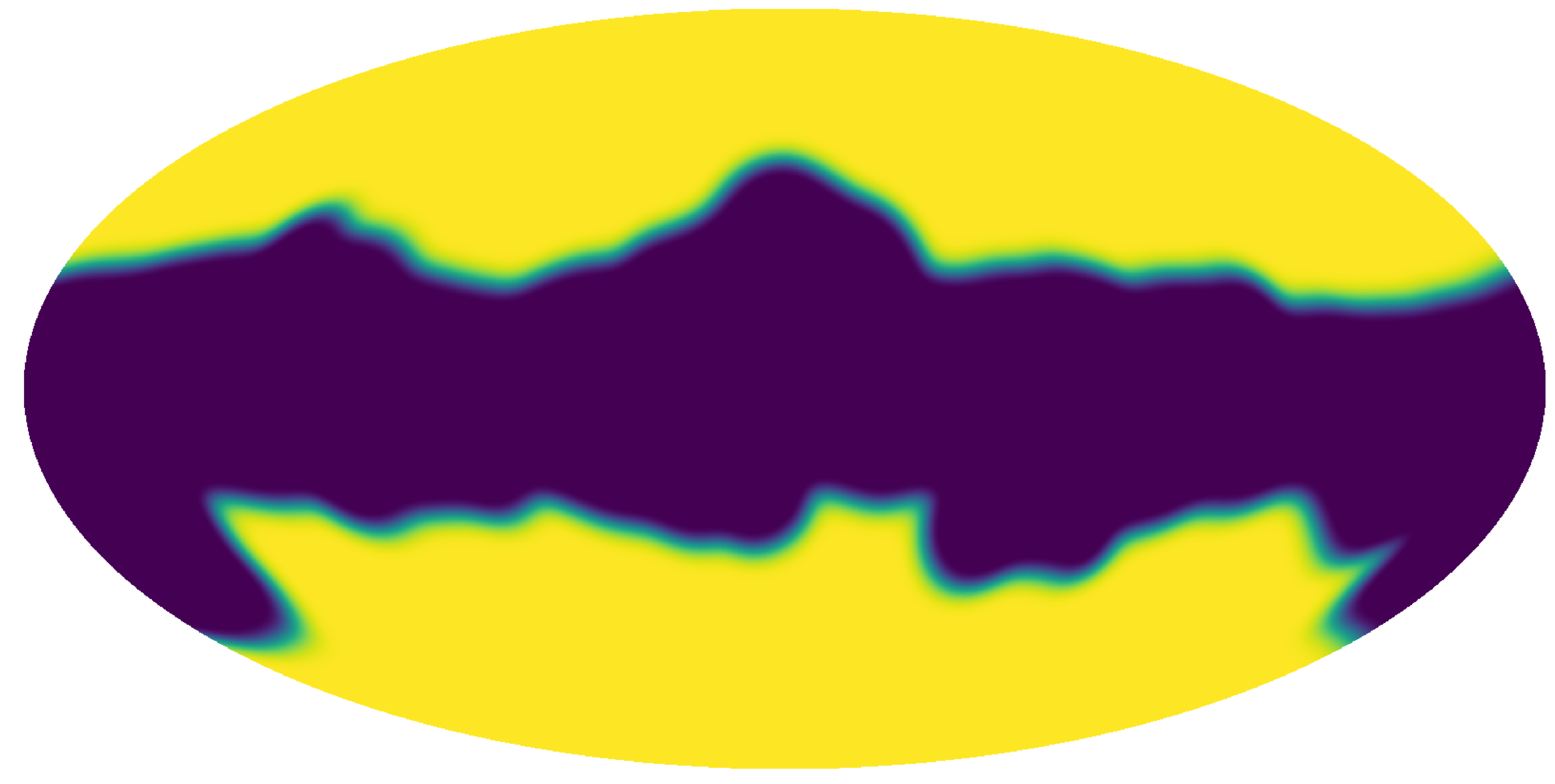}
        \caption{Planck's 60\,\% sky Galactic mask (``LB mask'')} \label{fig:planck_60pcsky_mask}
    \end{subfigure}
    \caption{Masks used in our analysis. They have been apodised using Gaussian kernels with a FWHM of $1^{\circ}$ in the case of (\subref{fig:so_bb_mask}) and $2^{\circ}$ in the case of (\subref{fig:planck_80pcsky_mask}) and (\subref{fig:planck_60pcsky_mask}).}
\end{figure}

In order to reduce the extent of mode coupling induced by masking and facilitate the deconvolution of mode-coupling matrices for power spectrum estimation, all masks are apodised. We obtain Galactic masks apodised with $2^{\circ}$ Gaussian kernels from the \href{http://pla.esac.esa.int/pla/#home}{Planck Legacy Archive}. On the other hand, the mask to be used for large-scale $B$-mode searches was kindly provided to us (in binary form) by the authors of~\citet{ref:SO_science_paper}. Apodisation of this mask is carried out using procedures built into the publicly-available code \texttt{NaMaster}~\citep{ref:namaster}, which ensures that regions originally excluded in the binary mask remain so after apodisation. Given that this mask has a relatively small footprint on the sky, we choose a FWHM of $1^{\circ}$ for the Gaussian smoothing kernel. The final mask is shown in figure~\ref{fig:so_bb_mask}. Henceforth, we refer to it as the ``SAT mask''. This will be the mask used to delineate the patch covered by a ground-based, small-aperture telescope going after large-angular-scale $B$-modes. We shall also consider the case of a space-based experiment targeting $B$-mode science, in the style of the upcoming LiteBird satellite~\citep{ref:litebird}. The sky patch covered by such an experiment is approximated by the 60\,\% sky Galactic mask from Planck shown in figure~\ref{fig:planck_60pcsky_mask}. We will sometimes refer to this as the ``LB patch''.

\begin{figure}
    \begin{subfigure}[t]{0.45\textwidth}
        \centering
        \includegraphics[width=\linewidth]{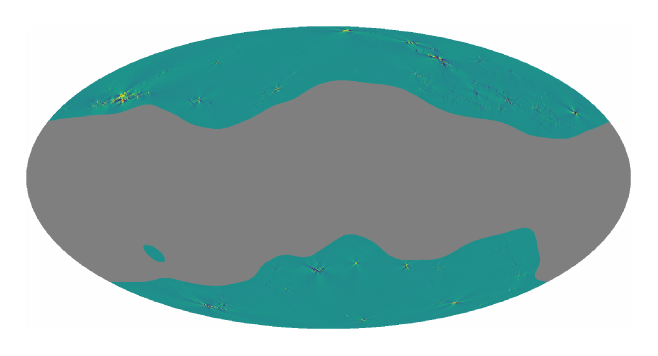}
        \caption{40\,\% unmasked} \label{fig:40pc_unmasked_template}
    \end{subfigure}
    \begin{subfigure}[t]{0.45\textwidth}
        \centering
        \includegraphics[width=\linewidth]{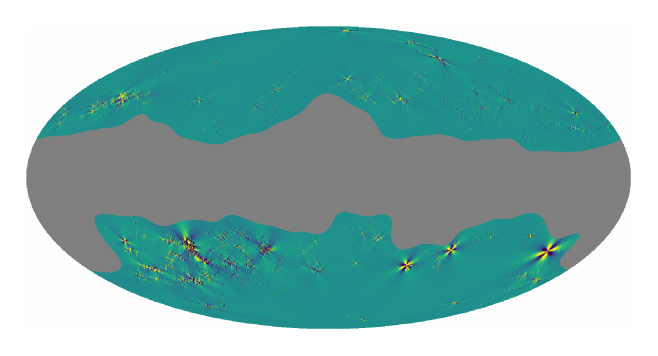}
        \caption{60\,\% unmasked} \label{fig:60pc_unmasked_template}
    \end{subfigure}
    \centering
        \begin{subfigure}[t]{0.45\textwidth}
        \centering
        \includegraphics[width=\linewidth]{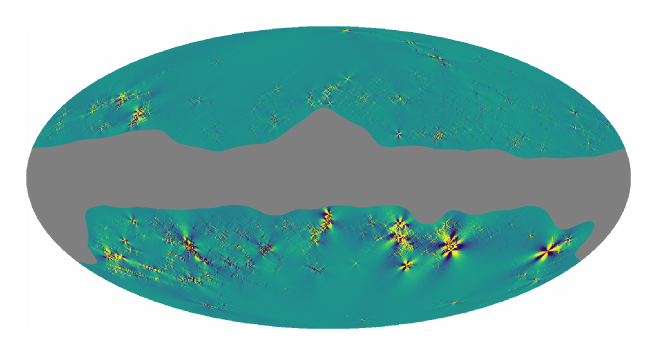}
        \caption{70\,\% unmasked} \label{fig:70pc_unmasked_template}
    \end{subfigure}
    \begin{subfigure}[t]{0.45\textwidth}
        \centering
        \includegraphics[width=\linewidth]{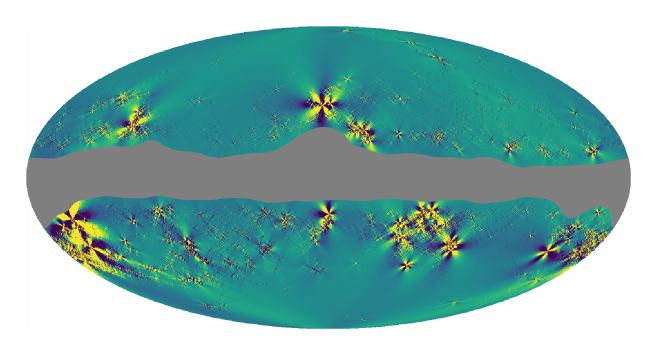}
        \caption{80\,\% unmasked} \label{fig:80pc_unmasked_template}
    \end{subfigure}
    \centering
        \begin{subfigure}[t]{0.45\textwidth}
        \centering
        \includegraphics[width=\linewidth]{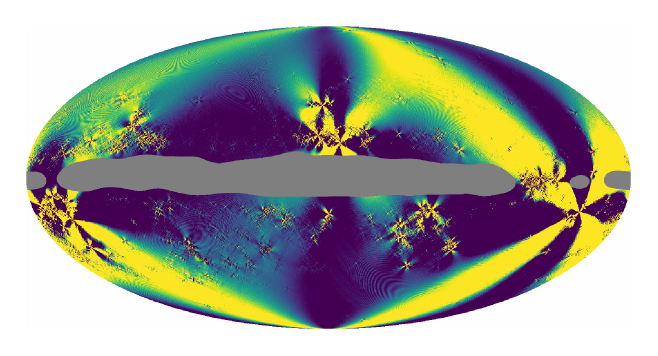}
        \caption{90\,\% unmasked} \label{fig:90pc_unmasked_template}
    \end{subfigure}
    \begin{subfigure}[t]{0.45\textwidth}
        \centering
        \includegraphics[width=\linewidth]{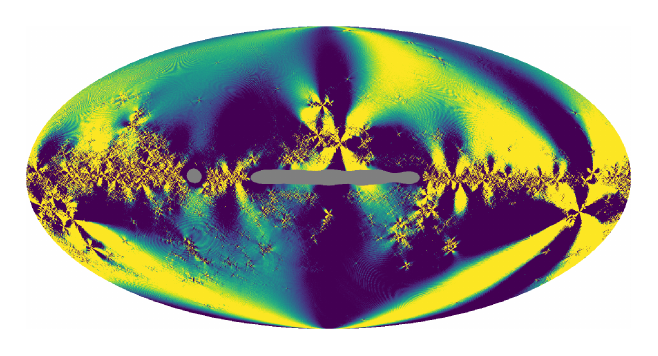}
        \caption{99\,\% unmasked} \label{fig:99pc_unmasked_template}
    \end{subfigure}

    \begin{subfigure}[t]{0.45\textwidth}
        \centering
        \includegraphics[width=0.5\textwidth]{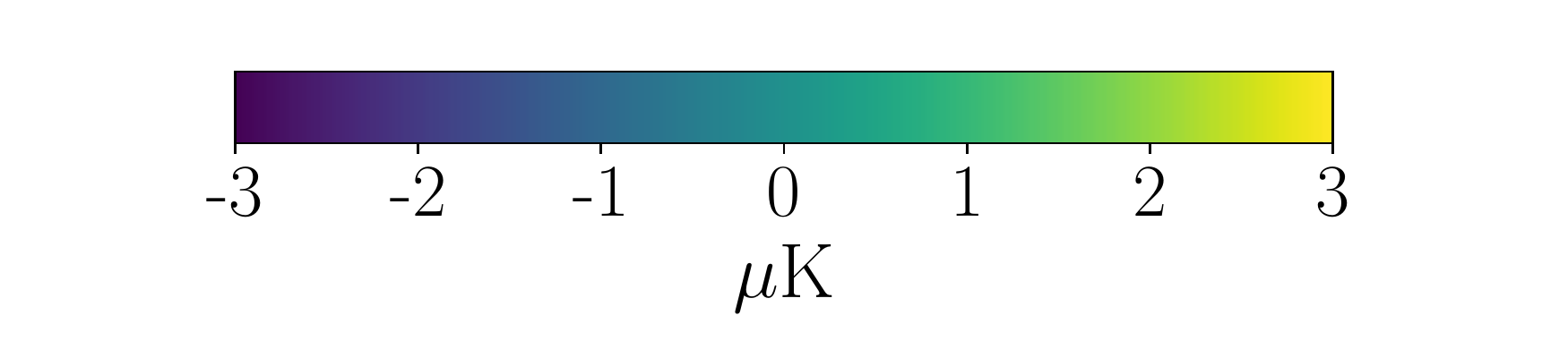}
    \end{subfigure}
    \caption{Stokes $Q$ parameter associated with $B$-mode templates generated from residual dust intensity and $E$-modes (as determined in section~\ref{sec:dust_transfer_funcs}), for different extents of Galactic masking applied to the input fields prior to Wiener filtering. Here, $X$\,\% unmasked is to be interpreted as meaning that $X$\,\% of the sky is left unmasked in each of the two fields, $E$ and $I$, from which the template is built (the masked region is shown in grey). The input dust $E$-modes are scaled to the residual level expected for standard ILC cleaning of the LAT in the \textit{goal} scenario, while dust total-intensity is scaled to levels appropriate for the residuals in the GNILC CIB maps over the SO SAT patch. Intensity modes below $l_{\mathrm{min}}=100$ are removed in all cases. The bright, hexadecapolar features are the expected consequence of leaving polarised clumps of Galactic emission unmasked in the inputs.}
    \label{fig:visualising_templates}
\end{figure}

\begin{figure}
        \begin{center}
        \includegraphics[width=0.6\columnwidth]{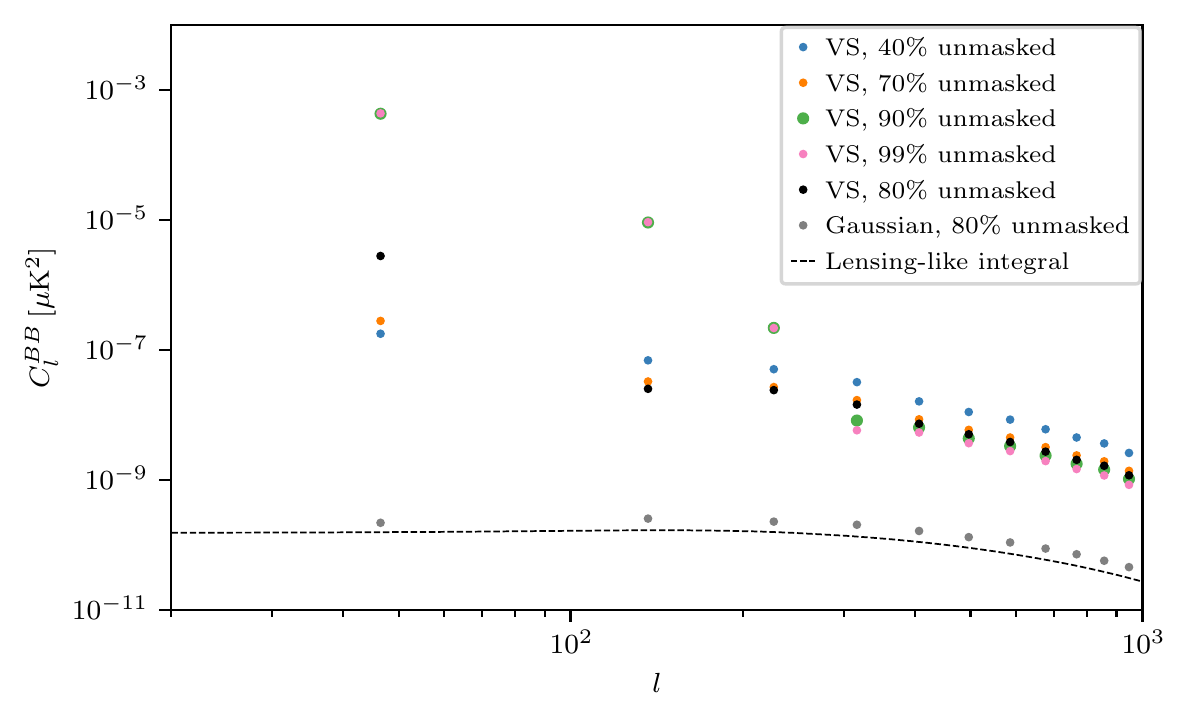}
        \end{center}
        \caption{Power spectra, measured on the SO SAT patch of figure~\ref{fig:so_bb_mask}, of the dust-only $B$-mode templates shown in figure~\ref{fig:visualising_templates}. We emphasise the comparison between the VS dust simulation and a Gaussian simulation (whose amplitudes in intensity and polarisation are made to match the power of the VS simulation measured on the SO SAT patch), after both are masked with the same 80\,\% sky mask of figure~\ref{fig:planck_80pcsky_mask} prior to Wiener filtering in the construction of the lensing template. We verify that the power in the Gaussian case is consistent with that expected from a leading-order lensing-like calculation based on the $E$ and $I$ dust spectra measured from the masked maps. On the other hand, the VS simulation displays more power, likely due to the relevance of the Galactic dust trispectrum. We also note the approximate convergence of template power between different masking fractions on intermediate and small scales. Power spectra, with the effect of the SO SAT mask deconvolved, are extracted with the code \texttt{NaMaster}, using bins with $\Delta l=90$.}
        \label{fig:impact_of_gal_mask}
\end{figure}

Though we expect the lensing $B$-mode template to be quite local -- owing to the fact that it is the product of predominantly intermediate and small-scale $E$-modes and lenses that give rise to lensed $B$-modes --- we saw in section~\ref{sec:possible_biases} that the delensing bias is sensitive to certain bispectra and trispectra. Since these scale as the third and fourth power of pixel values, they are very sensitive to bright structures in the dust residuals lying outside the footprint that we ultimately make $B$-mode measurements on. This is indeed what we see, already at the level of the template, when bright regions near the Galactic plane are left unmasked in the $E$ and $I$ fields from which these templates are constructed. Figure~\ref{fig:visualising_templates} shows several such templates for which input fields were first masked using Galactic masks of different extents before applying the Wiener filters in equation~(\ref{eqn:wiener_filters}). Bright artifacts outside the masked regions are readily apparent in those involving the largest unmasked fractions. Furthermore, these issues translate to increased power and scatter on the largest scales of the resulting templates, as shown by figure~\ref{fig:impact_of_gal_mask}.

These phenomena have the potential to hinder efforts to delens by accentuating the biases described in section~\ref{sec:possible_biases} and therefore should be mitigated by masking the Galactic region when constructing templates from wide-area data. We settle on the Planck mask that leaves 80\,\% of the sky available -- shown in figure~\ref{fig:planck_80pcsky_mask} -- as the option that minimises the masked fraction (to reduce mode-coupling effects in the construction of the Wiener-filtered fields) while keeping artifacts in the low-$l$ modes of the templates to acceptable levels. This scheme for masking fields going into templates will be implicit in all results quoted in the remainder of this paper. Of course, similar masking arrangements are inevitable in practical applications (at least from the ground) due to the telescopes' limited sky coverage. Even for space-based experiments, it will be important to mask away the galaxy -- and bright sources -- before constructing lensing templates.

To check that the power spectrum amplitudes we are seeing in figure~\ref{fig:impact_of_gal_mask} are physical in origin, we also show in the figure the results for a statistically-isotropic, Gaussian simulation. In this case, the $I$ and $E$ residual dust fields are simulated on the full sky from power spectra forced to match those of the VS simulation measured over the SO SAT patch. The full-sky Gaussian fields are masked with the 80\,\% sky mask prior to Wiener filtering in the construction of the lensing template. We verify that the amplitude resulting from the Gaussian simulation is consistent with an analytic calculation in the style of the leading-order lensing $B$-mode power spectrum where $C_l^{EE}$ and $C_l^{\kappa\kappa}$ are replaced with the dust spectra measured from the simulation. The amplitude in this Gaussian limit is significantly lower than that arising from the VS simulations. This is likely due to the power in the non-Gaussian VS simulations being boosted by the $\langle EIEI\rangle_{c}$ trispectrum of Galactic dust, though, in principle, statistical anisotropy could also be contributing to the difference seen between the VS simulation and the statistically-isotropic Gaussian simulations.

\subsection{Summary of simulation-based procedure for estimating delensing biases}\label{sec:ps}
We summarise here the procedure used to calculate the delensing biases of section~\ref{sec:possible_biases} from simulations. The steps are as follows:
    \begin{enumerate}
        \item Load in full-sky $I$, $Q$ and $U$ simulated dust maps at 353\,GHz. Take them to harmonic space to produce $I^{\mathrm{dust}}$, $E^{\mathrm{dust}}$ and $B^{\mathrm{dust}}$.
        \item Scale $E^{\mathrm{dust}}$ and $B^{\mathrm{dust}}$ to 145\,GHz as described in section~\ref{sec:dust_sims}.
        \item Apply the dust residual transfer functions measured in section~\ref{sec:dust_transfer_funcs} to each of $I^{\mathrm{dust}}$, $E^{\mathrm{dust}}$ and $B^{\mathrm{dust}}$.
        \item Simulate a CIB map, $I^{\mathrm{CIB}}$, in harmonic space using the method of section~\ref{sec:cib_sims}.
        \item Take $I^{\mathrm{dust}}$, $E^{\mathrm{dust}}$, $B^{\mathrm{dust}}$ and $I^{\mathrm{CIB}}$ to configuration space.
        \item Apply the apodised 80\,\% sky mask of figure~\ref{fig:planck_80pcsky_mask} to all of the maps.\label{step:apply_planck_mask}
        \item Take maps back to harmonic space.\label{step:processed_B_dust}
        \item Wiener filter $I^{\mathrm{CIB}}$ and $I^{\mathrm{dust}}$ according to equation~\eqref{eqn:wiener_filters}, with the total power spectrum in the denominator appropriate to the appropriate residual dust level under study. Mask modes below the $l_{\mathrm{min}}$ cutoff of choice.
        \item Using the methods described in appendix~\ref{sec:curved_sky_template}\footnote{These tools are packaged and made publicly-available under \texttt{https://github.com/abaleato/curved\_sky\_B\_template}}, build template $B$-modes of the form $\hat{B}\left(E^{\mathrm{dust}},I^{\mathrm{dust}}\right)$ and $\hat{B}\left(E^{\mathrm{dust}},I^{\mathrm{CIB}}\right)$.
        \item Convert these templates, as well as $B^{\mathrm{dust}}$ from step~\ref{step:processed_B_dust}, to $Q$ and $U$ maps.
        \item Apply the apodised SO SAT mask of figure~\ref{fig:so_bb_mask} to these $Q$ and $U$ maps.
        \item Use the pseudo-$C_l$ code \texttt{NaMaster} \citep{ref:namaster} to measure the correlations of interest (described in section~\ref{sec:possible_biases}) and deconvolve the mode couplings induced by the mask.\label{step:QU_to_EB_at_end}
        \item Fit a power-law to the bandpowers in the range $200<l<600$.

    \end{enumerate}{}
    Note that, in step~\ref{step:apply_planck_mask}, we use a Galactic mask from Planck rather than that from the SO LAT. There are two reasons for this: firstly, this mask is less restrictive and hence leads to less $E$-to-$B$ leakage in polarisation \citep{ref:lewis_eb_leakage, ref:bunn_eb_leakage}; and secondly, due to the local nature of the templates (see figure~\ref{fig:visualising_templates}) it should be possible to recover similar results if we were to restrict analysis to the LAT patch. In contrast, in step~\ref{step:QU_to_EB_at_end}, the SAT mask is used, which restricts analysis to a much smaller patch than that in step~\ref{step:apply_planck_mask}. However, there is no risk of $E$-to-$B$ leakage here since the polarisation maps do not contain $E$-modes by construction. For this reason, we do not bother to use the pure-$B$-mode formalism~\citep{ref:smith_06} when computing power spectra over the SO SAT patch.

    As a means of cross-validating our pipeline, we apply the same filtering and masking scheme to lensed simulations of CMB $E$-modes and combine these with $I^{\mathrm{CIB}}$ to build a lensing template. We verify that, when this template is used to delens $B$-modes, the theoretically-expected level of power (as determined by equation~\ref{eqn:delensed_power}) is obtained.

    Except where noted otherwise, we restrict the fields we work with to multipoles $8<l<2008$. This is the Planck GNILC bandwidth and, approximately, includes the full range of scales relevant for delensing the large-scale $B$-mode polarisation.

\section{Results}\label{sec:results}
\subsection{Bias from residual Galactic dust}\label{sec:bias_from_sims}
With the tools developed in the previous sections, we are now in a position to estimate the delensing bias arising from residual Galactic dust left over in the CMB and CIB maps involved in the delensing analysis. By applying the procedures of section~\ref{sec:ps} to dust simulations from VS, we shall see that there is indeed a bias dominated by the $BEI$ bispectrum of Galactic dust, that it is negative, and that it has a very red power spectrum, peaking on the largest angular scales.

\begin{figure}
    \begin{center}
    \includegraphics[width=0.85\columnwidth]{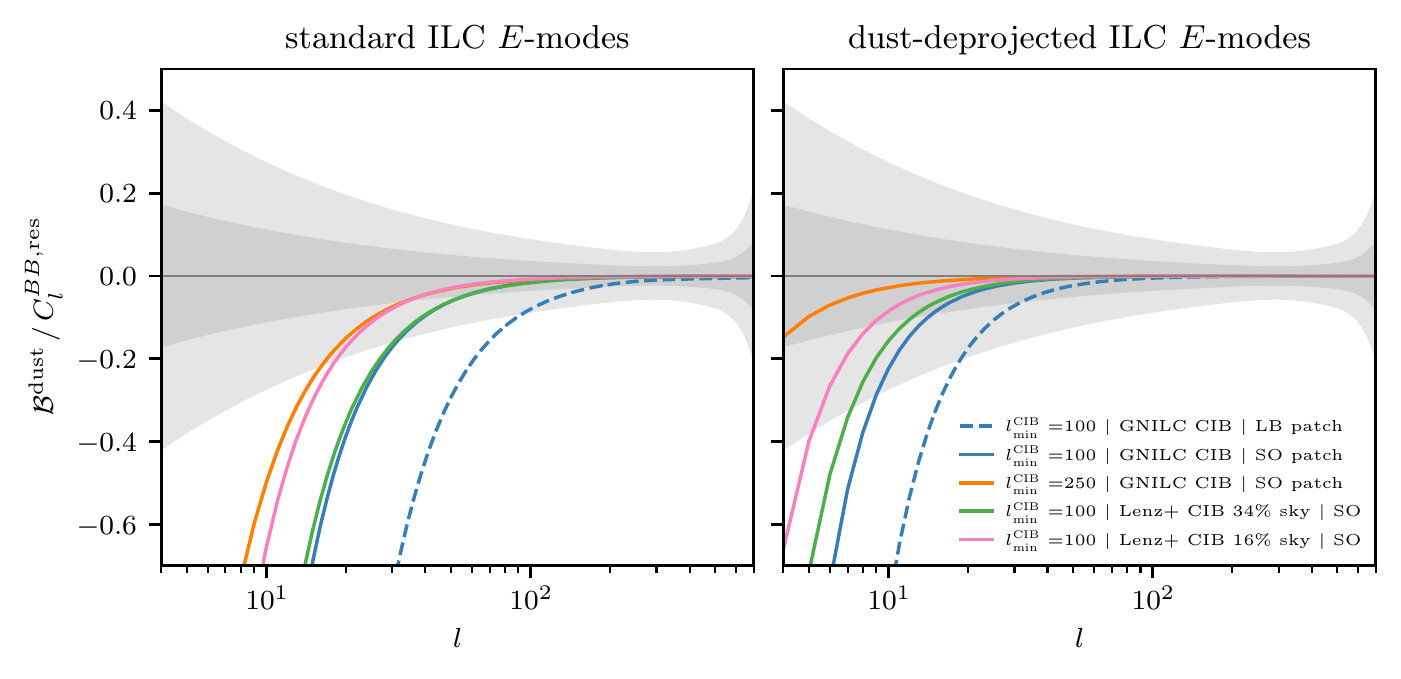}
    \end{center}
        \caption{Fractional bias on the angular power spectrum of residual lensing $B$-modes due to residual Galactic dust, plotted for several scenarios with differing residual dust levels. We model the residual lensing power as $C_l^{BB,\mathrm{res}}=0.7\tilde{C}_l^{BB}$, as expected after delensing with the CIB as the only tracer. \emph{Left panel:} $E$-mode residual for a standard ILC cleaning of SO's LAT (no dust deprojection and \emph{goal} noise levels assumed). \emph{Right panel:} $E$-mode residual as for SO's LAT (\emph{goal} noise levels), with deprojection of dust in the ILC cleaning. Foreground cleaning of $B$-modes is everywhere as for the SO SAT. Dashed lines are for measurements on the large `LB patch', while solid ones are on the smaller SO SAT patch. The shaded regions show the $\pm 1\,\sigma$ uncertainty on the power spectrum of delensed $B$-modes $C_l^{BB,\mathrm{del}} = 0.7\tilde{C}_l^{BB} + N_l^{BB}$, where $N_l^{BB}$ is the (\emph{goal}) noise power spectrum of the SO SAT. We calculate this uncertainty by assuming the delensed field is Gaussian, and that the experiment covers either $10\,\%$ (light grey) or $60\,\%$ of the sky (dark grey), and uses bins of $\Delta l=30$. These biases are dominated by the $BEI$ bispectrum of Galactic dust and are negative.}
    \label{fig:dust_bispectrum_bias}
\end{figure}

In figure~\ref{fig:dust_bispectrum_bias}, we show this power spectrum bias for several assumptions about the residual dust levels according to the type of CIB maps used, minimum-$l$-cuts applied to these, sky patch on which the $B$-mode measurement is carried out and whether or not dust is explicitly deprojected when ILC-cleaning the $E$-modes. We compare these biases to the power spectrum of residual lensing $B$-modes after delensing down to $70\%$ of the original lensing power  --- this is approximately the expected extent of delensing when using the GNILC CIB maps as the only tracer (c.f., figure~\ref{fig:delensing_efficiency_after_deprojecting}). For reference, we show via shaded regions the Gaussian part of the standard deviation of delensed $B$-modes when observed on either  $\sim 10\%$ or $\sim 60\%$ of the sky, representative of the SO $B$-mode survey patch or a larger patch observable from space, respectively. We model this delensed $B$-mode spectrum as comprising $70\%$ of the original lensing power and a white noise component in polarisation similar to that of the SO SAT's 145\,GHz channel in the \emph{goal} specifications (a beam $\theta_{\mathrm{FWHM}}$=17\,arcmin and a sensivity of $\Delta _{\mathrm{P}}=\sqrt{2}\times \,2.1\,\mu$K\,arcmin;~\citealt{ref:SO_science_paper}).

The very `red' spectral tilt displayed by the dust bias means that it will likely be smaller than the statistical uncertainty on the multipole range and sky regions probed by upcoming ground-based telescopes ($l\gtrsim 30$ and $f_{\mathrm{sky}}<0.1$), as long as modes of the CIB mass-tracer maps are removed for $l \lesssim 100$. From the left panel of figure~\ref{fig:dust_bispectrum_bias}, we see that this is true even in the case where dust is not deprojected as part of the foreground-cleaning procedure. However, one might want to be especially careful and work with dust-deprojected $E$-modes, as this is shown by the right panel of figure~\ref{fig:dust_bispectrum_bias} to mitigate the bias very significantly, while only marginally degrading delensing performance (at least for SO, see figure~\ref{fig:delensing_efficiency_after_deprojecting}).

On the other hand, the bias might pose challenges for space-based experiments (such as LiteBird) for two reasons: first, given their full-sky coverage and immunity to atmospheric noise, satellite missions will be able to probe polarisation down to the lowest multipoles, but it is on those largest scales that the bias is largest; and second, in observing a larger fraction of the sky, they encompass regions where the dust is brighter. These, compounded with the fact that larger sky coverage ($f_{\mathrm{sky}}\approx 0.6$) translates to lower statistical uncertainty, means that this bias could potentially confuse searches for primordial $B$-modes by space-based missions, unless their foreground cleaning capabilities are improved relative to SO. To illustrate this, we consider a scenario where the foreground cleaning is as for SO, but the analysis is performed on the larger `LiteBird' patch. Figure~\ref{fig:dust_bispectrum_bias} shows that the resulting bias is well in excess of the statistical errors, particularly when dust is not explicitly deprojected from the $E$-mode maps. The right panel of figure~\ref{fig:dust_bispectrum_bias} shows that doing this deprojection is very effective at reducing the amplitude of the bias, but even then it is large enough to signficantly compromise contraints on $r$ (as we will quantify shortly). Fortunately, space-based experiments targeting large-scale $B$-mode science will take advantage of their privileged vantage point above the atmosphere to observe in many more frequency channels channels than SO -- LiteBird, for instance, is expected to have 15 frequency bands~\citep{ref:litebird}. Hence, foreground cleaning of, at least, the large-scale $B$-modes should be more extensive than what is assumed in the figure, which has dust residuals appropriate for SO. Since the bias is dominated by the $BEI$ bispectrum of dust, its amplitude is very sensitive to improved cleaning of the CMB polarisation fields.

For a fixed level of foreground residuals, the bias can be mitigated by filtering out the largest scales of the CIB maps, which tend to be relatively more contaminated by dust (this is particularly acute for the GNILC products, as discussed in section~\ref{sec:dust_transfer_funcs}). In all likelihood, upcoming implementations of CIB delensing will be complemented with internal reconstructions, which are able to reconstruct the largest-angular-scale lenses very accurately. Even if the largest scales are discarded altogether, the impact is expected to be small since it is scales smaller than that (chiefly $200<l<500$) that are most informative for delensing: for instance, \citet{ref:sherwin_2015} show that delensing efficiency only worsens by $12\,\%$ if the CIB fields are high-pass filtered with $l_{\mathrm{min}}=150$ instead of $l_{\mathrm{min}}=60$. Consequently, the recommendation we issue here of removing those largest scales from the CIB maps should come at very little cost in terms of delensing efficiency.

\begin{figure}
    \begin{center}
    \includegraphics[width=0.85\columnwidth]{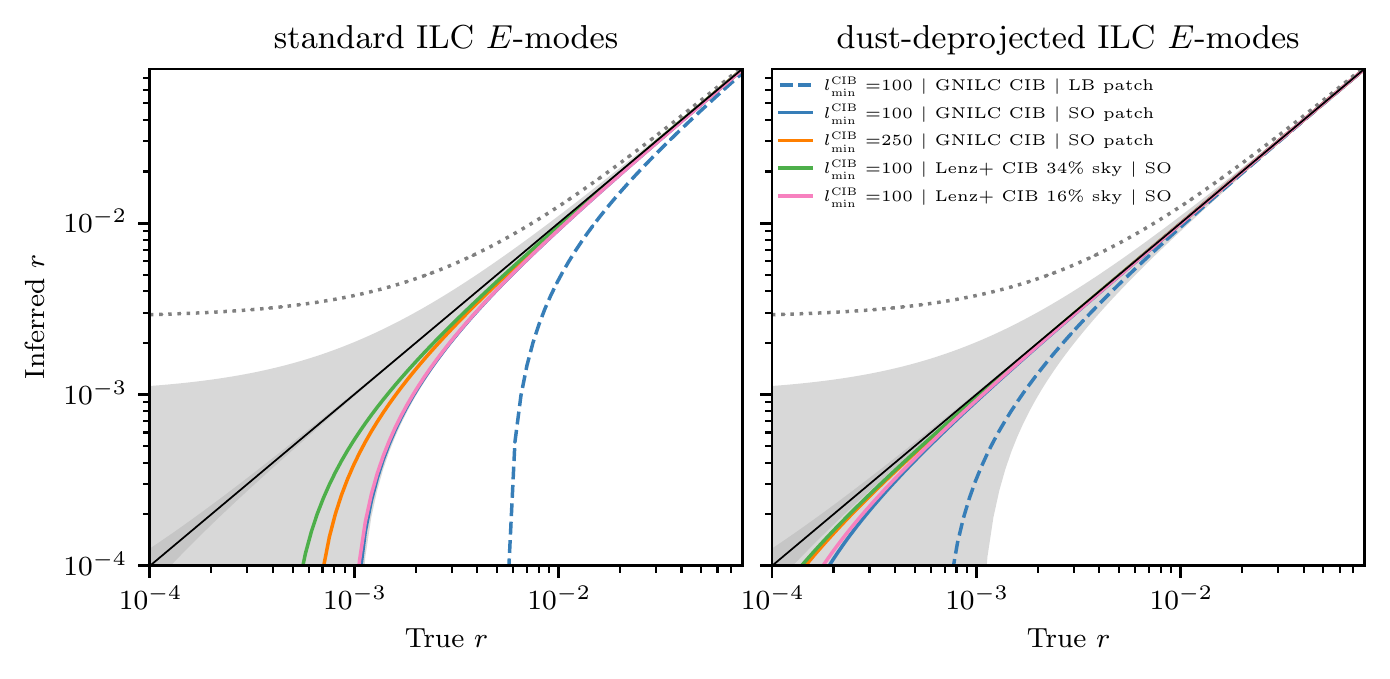}
    \end{center}
        \caption{Bias on the inferred tensor-to-scalar ratio due to unmodelled contributions from residual non-Gaussian Galactic dust to the power spectrum of $B$-modes delensed with the CIB. Dashed lines are for measurements on the `LB patch' (approximately $60\,\%$ of the sky), and solid ones are on the smaller SO SAT patch (approximately $10\,\%$ of the sky). Foreground cleaning of $B$-modes is everywhere as for the SO SAT.  Dust residuals in the $E$-modes are as for a harmonic ILC cleaning of the SO LAT at \emph{goal} noise levels, either in a standard implementation (left panel) or after deprojection of Galactic dust (right panel). We constrain $r$ using scales $l_{\mathrm{min}}=20$ and $l_{\mathrm{max}}=200$ or $l_{\mathrm{min}}=2$ and $l_{\mathrm{max}}=200$ when working with either the SO SAT patch or the `LB' patch, respectively. For reference, the shaded regions show the approximate $\pm1\,\sigma$ uncertainty for $r=0$ of experiments covering either $10\,\%$ (light grey) or $60\,\%$ (dark grey) of the sky with the noise level of the SO SAT, no foregrounds, delensing down to $70\,\%$ of the original lensing $BB$ power and when $r$ is constrained over the multipole ranges described above for each of the sky fractions. For comparison also, the dotted lines show the size of the bias on $r$ if residual dust $B$-modes in the SO SAT maps are as in figure~\ref{fig:jpl_vs_gnilc_dust_res} and are not modelled in the $BB$ power spectrum.}
    \label{fig:r_bias}
\end{figure}

In figure~\ref{fig:r_bias}, we quantify how the power spectrum biases of figure~\ref{fig:dust_bispectrum_bias} propagate to biases on the inferred value of the tensor-to-scalar ratio, $r$. We use (e.g.,~\citealt{ref:roy_et_al_21})
\begin{equation}
    \Delta \hat{r} = \left( \sum_{l=l_{\mathrm{min}}}^{l_{\mathrm{max}}} \left[C_l^{BB,\mathrm{prim}}(r=1)\right]^2 / \mathrm{Var}\left(C_l^{BB,\mathrm{del}}\right) \right)^{-1}  \sum_{l=l_{\mathrm{min}}}^{l_{\mathrm{max}}} C_l^{BB,\mathrm{unmodelled}} C_l^{BB,\mathrm{prim}}(r=1) / \mathrm{Var}\left(C_l^{BB,\mathrm{del}}\right)
\end{equation}
to translate an improper modelling of the $B$-mode power spectrum to a shift in $r$. Here,
$C_l^{BB,\mathrm{prim}}(r=1)$ is the angular power spectrum of primordial $B$-modes with $r=1$, $\mathrm{Var}(C_l^{BB,\mathrm{del}})$ is the variance of the power spectrum of delensed $B$-modes and $C_l^{BB,\mathrm{unmodelled}}$ is the part of the delensed $B$-mode spectrum that we have failed to model. We compute $\mathrm{Var}(C_l^{BB,\mathrm{del}})$ assuming the delensed $B$-modes to be Gaussian, and to feature only experimental noise (at a level appropriate for the SO SAT's 145\,GHz \emph{goal} specifications), primordial signal and $70\,\%$ of the original lensing power. For constraints on the SO SAT patch, we use $l_{\mathrm{min}}=20$ and $l_{\mathrm{max}}=200$, and for the larger `LB' patch we set $l_{\mathrm{min}}=2$ and $l_{\mathrm{max}}=200$. We compare the estimated shifts to the standard deviation on inferences of $r$ in the limit $r=0$ expected of experiments covering either $10\,\%$ or $60\,\%$ of the sky, with the noise levels of the SO SAT, and in the limit of no foreground $BB$ power and residual lensing power of $70\,\%$ of the original. Figure~\ref{fig:r_bias} shows that the bias discussed here results in inferences of $r$ that are systematically lower that the truth, although for ground-based experiments this is still within the bounds of statistical errors. This is in contrast with the impact of failing to model dust residuals on large angular scales of the $B$-mode power spectrum, in which case one tends to over-estimate the true value of $r$. As we show in the figure, failing to model dust $B$-mode residuals with a level of power as estimated in figure~\ref{fig:jpl_vs_gnilc_dust_res} for the SO SAT could result in a bias larger than those considered in this section, but in the opposite direction.

By combining the mitigation techniques described here, we ultimately expect the delensing bias from residual Galactic dust to be contained to levels where it does not compromise constraints on $r$, for virtually any application of CIB delensing. In particular, deprojection of dust from the $E$-modes proves to be particularly effective, while incurring only a small degradation in delensing efficiency, as shown in figure~\ref{fig:delensing_efficiency_after_deprojecting}. However, this work highlights that, if we are to exploit the full power of the CIB for delensing, realistic simulations of Galactic dust, extending to small angular scales, ought to be developed so that they can be used to model and properly account for the expected amplitude and shape of the bias.

\subsection{Intrinsic bias from non-Gaussianity of the CIB}\label{sec:analytic_cib_bias_calc}
Emission from UV-heated dust in star-forming galaxies -- the diffuse component of which constitutes the CIB -- is linearly polarised. This is
thought to arise from the alignment of asymmetric dust grains in the inter-stellar medium of these galaxies with the local magnetic field direction. For observations along a particular line of sight, the stacking of emission from a large number of sources, each with incoherent, randomly-orientated polarisation directions, gives rise to depolarisation of the observed, integrated radiation.
This depolarisation is practically complete for low-flux sources, of which there are very many. On the other hand, the number of bright sources is more modest, and their incomplete averaging gives rise to a net polarisation of the observed light. Experimental applications are able to mitigate this effect somewhat by removing the bright sources that they are able to resolve (e.g.,~\citealt{ref:planck_point_sources}), but the contribution from those that are not resolved ought to be modelled.

We determined in section~\ref{sec:possible_biases} that, if the CIB itself has non-vanishing $\langle BEI\rangle_c$ and $\langle EIEI\rangle_{c}$ correlations, the power spectrum of delensed $B$-modes will be biased by the following two terms:
\begin{align}{}\label{eqn:cib_bispectrum_bias_firstoccurence}
\mathcal{B}^{BEI}_l = \frac{4}{(2\pi)^2} \int & \frac{d^2\bmath{l}'}{(2\pi)^2} \frac{\bmath{l}'\cdot (\bmath{l}-\bmath{l}')}{|\bmath{l}-\bmath{l}'|^2} \mathcal{W}^{E}_{l'}\mathcal{W}^{I}_{|\bmath{l}-\bmath{l}'|} \sin{2(\psi_{\bmath{l}'}-\psi_{\bmath{l}})} \langle B(-\bmath{l}) E(\bmath{l}') I(\bmath{l}-\bmath{l}')\rangle' ;
\end{align}
and
\begin{align}{}\label{eqn:cib_trispectrum_bias_general}
\mathcal{B}^{EIEI}_l = - \frac{4}{(2\pi)^2}
\int & \frac{d^2\bmath{l}'d^2\bmath{l}''}{(2\pi)^4} \frac{\bmath{l}'\cdot (\bmath{l}-\bmath{l}')}{|\bmath{l}-\bmath{l}'|^2}
\frac{\bmath{l}''\cdot (\bmath{l}+\bmath{l}'')}{|\bmath{l}+\bmath{l}''|^2}\mathcal{W}^{E}_{l'}\mathcal{W}^{E}_{l''}\mathcal{W}^{I}_{|\bmath{l}-\bmath{l}'|}\mathcal{W}^{I}_{|\bmath{l}+\bmath{l}''|} \sin{2(\psi_{\bmath{l}'}-\psi_{\bmath{l}})}\sin{2(\psi_{\bmath{l}''}-\psi_{\bmath{l}})}\nonumber \\ & \times \langle E(\bmath{l}')
I(\bmath{l}-\bmath{l}') E(\bmath{l}'') I(-\bmath{l}-\bmath{l}'')\rangle'_{c} \, ,
\end{align}
where the primes on the expectation values denote that the delta-function has been dropped.

In order to obtain an analytic understanding of these biases and be able to calculate higher-point functions of the CIB, we introduce a minimal model where the source polarisation fraction is constant and the polarisation angles of sources are uncorrelated. A detailed derivation of the $BEI$ bispectrum of the CIB in this limit is provided in appendix~\ref{appendix:BET_bispectrum}. Following equation~\eqref{eqn:cib_bispectrum_general}, we learn that this sources a bias given in CMB temperature units by
\begin{align}\label{eqn:cib_bispectrum_bias}
\mathcal{B}^{BEI}_l = 2p^2 G_{353\,\mathrm{GHz}}G^2_{145\,\mathrm{GHz}} \int & \frac{d^2\bmath{l}'}{(2\pi)^2} \frac{\bmath{l}'\cdot (\bmath{l}-\bmath{l}')}{|\bmath{l}-\bmath{l}'|^2} \mathcal{W}^{E}_{l'}\mathcal{W}^{I}_{|\bmath{l}-\bmath{l}'|} \sin^2{2(\psi_{\bmath{l}'}-\psi_{\bmath{l}})} \nonumber \\
&\times \int dz \left(\frac{I^{\mathrm{CIB}}[145(1+z)\,\mathrm{GHz}]}{I^{\mathrm{CIB}}[353(1+z)\,\mathrm{GHz}]}\right)^2 \left[S_{353\,\mathrm{GHz}}^{(3)}(z) + \frac{H(z)}{c r^2(z)}S_{353\,\mathrm{GHz}}^{(2)}(z)S_{353\,\mathrm{GHz}}^{(1)}(z)P_{g}\left(|\bmath{l}-\bmath{l}'|/r(z);z\right)\right],
\end{align}
where $I^{\mathrm{CIB}}(\nu)$ is the CIB SED at frequency $\nu$, $G_\nu$ is a conversion factor\footnote{For narrow frequency bands, this can be calculated using, for example, equation~(8) of~\citet{ref:planck_HFI_spectral_response}.} to go from specific intensity to differential CMB temperature units at frequency $\nu$, $p$ is the polarisation fraction of CIB galaxies, which we assume to have a constant value of $5\,\%$\footnote{This value is more conservative than what has recently been discussed in the literature: \citet{ref:lagache_19}, for example, suggest a smaller $p\approx1\,\%$. The rationale behind such low figures is that the complex arrangement of magnetic field lines in dusty, star-forming galaxies is expected to bring about extensive depolarisation when integrated over the whole galaxy.}, and $P_{g}(k;z)$ is the galaxy power spectrum. We have also defined
\begin{equation}\label{eqn:Sn_definition}
    S^{(n)}_{\nu}(z) \equiv \int^{s_{\mathrm{max}}}_{s_{\mathrm{min}}}dS_\nu S_\nu^n \frac{dN}{dS_\nu dz d\Omega},
\end{equation}
that is, an integral over the $n$th power of source flux density, $S_\nu$, weighted by $dN/dS_\nu dz d\Omega$ -- the number of sources per steradian, at a given redshift in a certain flux-density range $[s_{\mathrm{min}}, s_{\mathrm{max}}]$ at frequency $\nu$. These number counts can be obtained by fitting a model to sub-mm data. In this work, we use those of~\citet{ref:bethermin_counts}.

Since we have assumed that the polarisation angles of different galaxies are uncorrelated, the two polarisation legs of the bispectrum must be sourced by the same galaxy, so there can be no 3-source contribution. Hence, both terms in equation~\eqref{eqn:cib_bispectrum_bias} can be interpreted as shot-noise contributions where the same galaxy appears in more than one leg -- the first term comes from a single galaxy and the second, from two. If a mechanism exists for aligning galaxy spins (and, hence, polarisations) in the filamentary cosmic web (see, e.g., \citealt{ref:piras_18} and \citealt{ref:codis_18}), then there can, in principle, also be a 3-source contribution dependent on the 3D galaxy bispectrum. However, such a mechanism has been shown by \citet{ref:feng_holder} to produce contributions to the CIB power spectrum that are several orders of magnitude smaller than the shot noise, so we will be ignoring it henceforth.

In order to evaluate equation~\eqref{eqn:cib_bispectrum_bias}, we need to compute the power spectrum of the galaxies comprising the CIB emission. Galaxies are known to be biased tracers of the matter distribution~\citep{ref:kaiser_84}, so additional prescriptions are needed to relate their clustering statistics to those of the matter. This could be done, for instance, by applying the tools of the halo model to the CIB, as was done by~\citet{ref:lacasa_halo_model}. However, given the number of approximations we have already made, we pursue a cruder approach and relate the power spectrum of galaxy fluctuations to that of the matter by means of an effective galaxy bias $b_{g}(k,z)$, relating the over-densities of galaxy number density and matter as $\delta_{g}(k,z)=b_{g}(k,z)\delta_{m}(k,z)$. We use the value measured by~\citet{ref:planck_13_cib}, which assumes that the galaxy bias is scale-independent. If, as evidence suggests, galaxy formation is chiefly a local process, then this ought to be a very good approximation on large scales~\citep{ref:coles_93}, where the two-halo term dominates the galaxy power spectrum. Those are precisely the scales of the power spectrum that we probe in the bispectrum configuration we are dealing with here. To see why, notice that, given $\mathcal{W}^{E}_l$ decreases quickly below $l>2000$ due to limited sensitivity, the multipole in the argument of the galaxy power spectrum in equation~\eqref{eqn:cib_bispectrum_bias} is at most that large, which in turns forces the wavenumber $k=l/r(z)$ (where $r(z)$ is the comoving distance out to a redshift of a few, i.e., a few $1000\,\text{Mpc}$) to be small enough that we are probing a regime where the linear bias approximation roughly holds.

\begin{figure}
    \begin{center}
    \includegraphics[width=0.6\columnwidth]{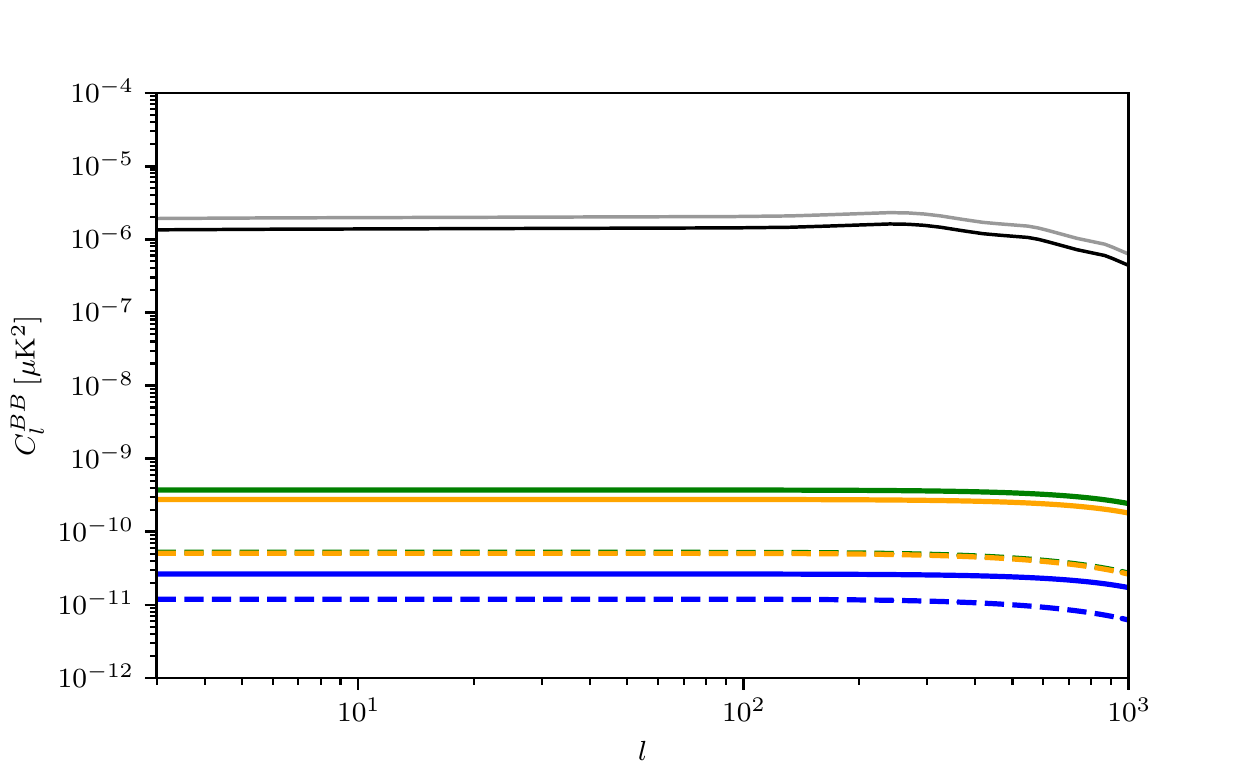}
    \end{center}
    \caption{Delensing biases from angular bispectra of the CIB in our minimal model with uncorrelated source polarisation angles. Solid (dashed) lines show the one(two)-source contribution from different point-source masking schemes with $s_{\mathrm{max}}=\{200,20,2\}$\,mJy (in green, orange and blue, respectively). We evaluate our predictions using $0.2<z<6$, $s_{\mathrm{min}}=10^{-5}$\,mJy, a fixed galaxy polarisation fraction of $5\,\%$, and the number counts of~\citet{ref:bethermin_counts}. For comparison, we show also the fiducial power spectrum of lensed $B$-modes before (grey) and after (black) delensing with the CIB such that the power is reduced by 30\,\%. We note that the two-source contribution is virtually unchanged if the flux cut is lowered from 200\,mJy to 20\,mJy, so the orange and green dashed lines lie on top of each other.
    }
    \label{fig:cib_biases}
\end{figure}

Finally, we evaluate the delensing bias of equation~\eqref{eqn:cib_bispectrum_bias} for different flux cuts typical of current and upcoming experiments, and show the results in figure~\ref{fig:cib_biases}. We see that the one-source contribution dominates over the two-source, but it is everywhere negligible compared to the amplitude of the delensed $B$-mode spectrum. This is the case even though we have assumed that, beyond masking the brightest sources, no attempt is made to mitigate the CIB contribution to the CMB polarisation maps using, for example, multi-frequency cleaning techniques.

On the other hand, the bias of equation~\eqref{eqn:cib_trispectrum_bias_general}, sourced by the $EIEI$ trispectrum of the CIB, can, in principle, receive contributions of one-, two- and three-source type; where in all cases, the two polarisation modes are sourced by the same galaxy. As before, these can be evaluated in the framework of the halo model, this time requiring computation of the 3D halo bispectrum to obtain the three-source term. However, we argue that the trispectrum ought to be subdominant to the bispectrum (which we have shown is negligible) for the level of point-source removal expected of upcoming experiments, since the former scales as the fourth power of the flux density while the latter goes as the third. In appendix~\ref{appendix:toy_model_cib_higherpointfns}, we motivate this claim by comparing the 1-source contributions to both higher-point functions, showing that, indeed, the relevant trispectrum is smaller than the bispectrum in this even simpler model (see figure~\ref{fig:cib_biases_discrete}).

Our conclusion is that biases sourced by higher-point functions of the CIB are negligible for the levels of point-source masking and delensing expected of any foreseen applications of the CIB-based method considered in this paper.

\section{Validation}\label{sec:validation}
\begin{figure}
        \begin{center}
        \includegraphics[width=0.6\columnwidth]{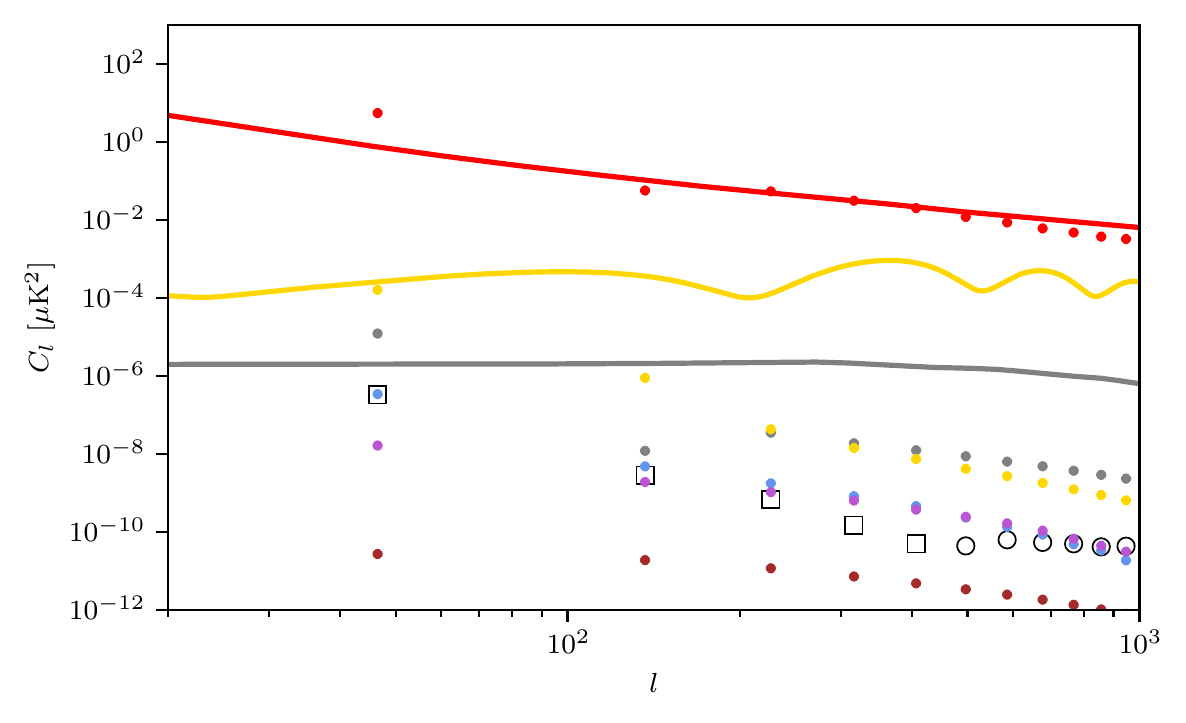}
        \end{center}
        \caption{Comparison of power spectrum amplitudes and biases for a number of fields involved in the delensing procedure. \emph{Red dots}: residual dust intensity after filtering the VS simulation at 353\,GHz with the GNILC CIB residual transfer function of figure~\ref{fig:jpl_vs_gnilc_dust_res}. \emph{Red line}: theoretical CIB intensity at 353\,GHz. \emph{Yellow dots}: VS dust $E$-mode power at 145\,GHz scaled to the residual dust amplitude expected for the SO LAT after ILC-cleaning with dust deprojection. \emph{Yellow line}: theoretical, lensed CMB $E$-mode. \emph{Grey dots}: VS dust $B$-mode power at 145\,GHz, scaled to the residual dust amplitude expected for the SO SAT. \emph{Grey line}: theoretical, lensed CMB $B$-modes. \emph{Purple dots}: power spectrum of a $\hat{B}(E^{\mathrm{dust}},I^{\mathrm{dust}})$ template constructed with $E^{\mathrm{dust}}$ and $I^{\mathrm{dust}}$ with amplitudes as shown in this plot. \emph{Brown dots}: power spectrum of a $\hat{B}(E^{\mathrm{dust}},I^{\mathrm{CIB}})$ template. \emph{Blue dots}: delensing bias associated with the $\langle B^{\mathrm{dust}}E^{\mathrm{dust}}I^{\mathrm{dust}}\rangle_{c}$ bispectrum for residual levels as shown in this plot (note that this bias is negative, though here we plot it as a positive quantity). The blue, brown and purple points, along with the terms evaluated in figure~\ref{fig:cib_biases}, comprise all the contributions to equation~\eqref{eqn:bias_def} that we identified as being potentially relevant. Comparison of each of these with the total delensing bias --- shown as empty squares where negative, and empty circles where positive --- shows that the dust bispectrum term dominates on the scales relevant for primordial $B$-mode searches. All empirical measurements are carried out on the LB patch, using \texttt{NaMaster} and a binning scheme with $\Delta l =90$.}
        \label{fig:compare_all_spectra}
\end{figure}

In this section, we aim to validate the results of section~\ref{sec:bias_from_sims}, where we used simulations to quantify the delensing bias associated with residual Galactic dust. For reference, the power spectra of all the fields involved in the delensing procedure are shown in figure~\ref{fig:compare_all_spectra}.

Of course, the results of section~\ref{sec:bias_from_sims} are only valid in as much as the simulations on which they rely faithfully reproduce the true behaviour of dust on the sky. Given the dearth of small-scale polarisation data at the time of writing, it is not possible to gauge the accuracy of simulations by comparing them to observations. We can, however, test the results in two critical ways: firstly, we can ensure that the delensing bias (which, as we have already established, is dominated by the $\langle BEI\rangle_{c}$ bispectrum of dust) does not appear when our pipeline is applied to Gaussian simulations with the same power spectra; and secondly, we can verify that the biases that ensue after either dust data from Planck or the VS dust simulations are used in the analysis display similar characteristics after both inputs are restricted to the same scales.

\subsection{Validation against Gaussian simulations}
In order to guarantee that the delensing bias we identified in section~\ref{sec:bias_from_sims} -- associated, recall, with the $\langle BEI\rangle_{c}$ bispectrum of Galactic dust -- is physical in origin and not some artifact of our analysis, we re-run on Gaussian simulations of Galactic dust with the same power spectra as their non-Gaussian counterparts from VS.

We work with full-sky simulations of Galactic dust at 145 GHz. First, we fit smooth curves to the ratios of $TT$, $EE$ and $BB$ power in these simulations to those measured in the VS simulations on scales $60<l<2000$, after the latter have been scaled to the appropriate frequency and residual amplitude, and restricted to the region allowed by the SAT mask of figure~\ref{fig:so_bb_mask}. The square root of these smooth transfer functions are then used to filter the Gaussian simulations, obtaining Gaussian maps with the same power spectra as the VS simulations.

\begin{figure}
        \begin{center}
        \includegraphics[width=0.8\columnwidth]{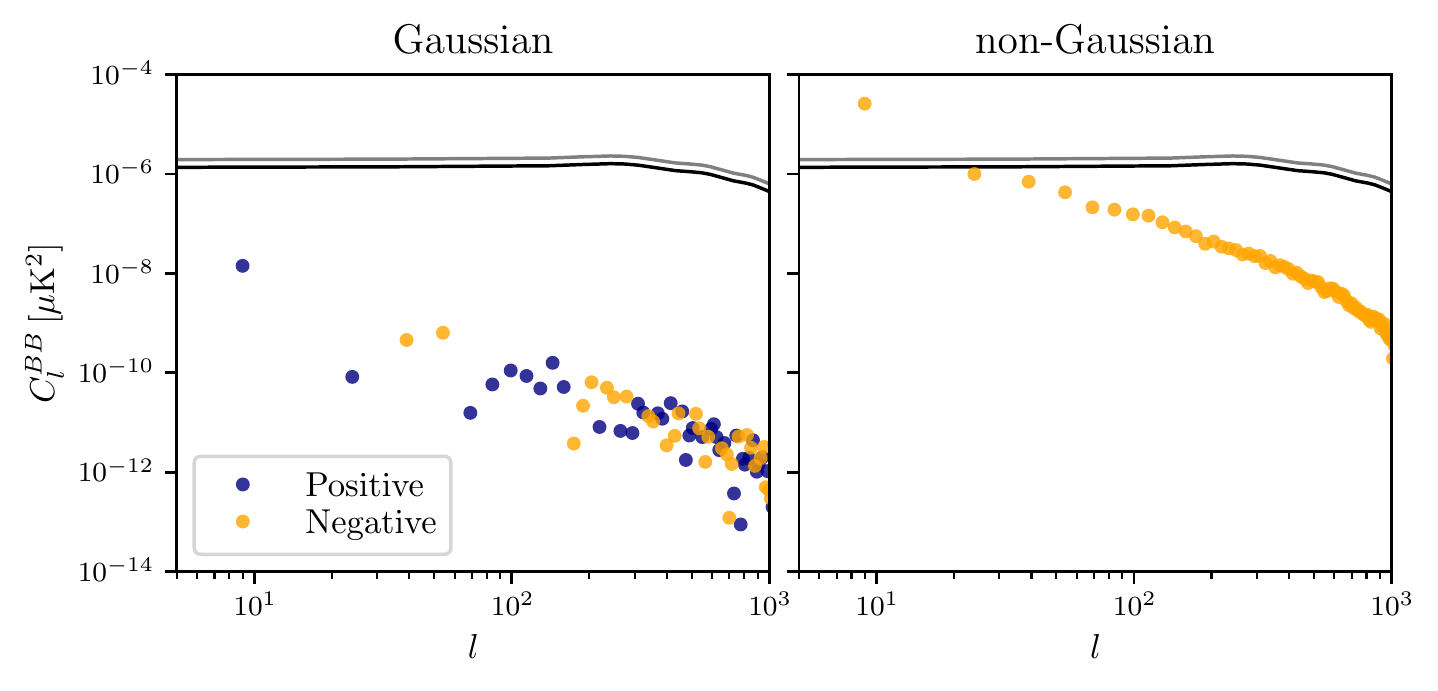}
        \end{center}
        \caption{Delensing bias associated with the bispectrum of Galactic dust as calculated from a single realisation of the VS simulation (right panel), compared to that measured from a Gaussian simulation of the dust (left panel) with the same power spectra. The same transfer functions are applied in both cases to induce the scale-dependent dust residuals estimated in section~\ref{sec:dust_transfer_funcs}. Spectra are measured on the ``LB patch'' shown in figure~\ref{fig:planck_60pcsky_mask} and binned with $\Delta l=15$. For comparison, the amplitude of the lensing $B$-mode spectrum is plotted before (grey) and after (black) delensing using the CIB as the only matter tracer. While the measurement on Gaussian simulations is consistent with scatter around zero, the non-Gaussian case shows a distinct bias-like behaviour.}
        \label{fig:validate_on_gaussian}
    \end{figure}

Next, we use these re-scaled Gaussian simulations in the analysis steps described in section~\ref{sec:ps}. As shown in figure~\ref{fig:validate_on_gaussian}, the output is consistent with pure scatter, in contrast with the distinct bias arising from the non-Gaussian simulations.

\subsection{Validation against Planck data}
Given that the results of section~\ref{sec:bias_from_sims} are only as accurate as the simulations on which they rely, we set out now to validate those findings against Planck data on the scales where the latter has sufficient signal-to-noise. More specifically, we compare the output of our analysis procedure (detailed in section~\ref{sec:ps}) when applied to the VS simulations to the case where the temperature and polarisation maps we use are instead the Planck 2018 full-mission or GNILC products, both at 353\,GHz. (The Planck products we use are available on the \href{http://pla.esac.esa.int/pla/#home}{Planck Legacy Archive}.)

The idea is that the dust component of the observations is to be processed in the same ways as the dust in the simulations. For this reason, we apply the same filtering protocol to both, aiming to retrieve the same residual dust fractions. In addition to this, we extrapolate the polarisation observations at 353\,GHz to 145\,GHz with an SED appropriate for dust, to estimate the dust at this frequency before foreground cleaning. This is done by assuming that the dust behaves as a modified blackbody with parameters given in table~\ref{tab:constants} and applying the relevant wide-band unit-conversions for the 143 and 353\,GHz channels of Planck. Furthermore, specific intensity measurements made by broad-band instruments are necessarily related to a reference frequency and calibration SED profile. In the Planck 353\,GHz band, this is the CMB dipole. In order to relate measurements on that band to a thermal dust SED, we apply a multiplicative colour correction as given by~\citet{ref:planck_18_polarised_dust}. An explanation of the philosophy of such corrections can be found in~\citet{ref:planck_HFI_spectral_response}.

As is the case with all astronomical observations, the Planck data are convolved with the instrument's beam. In order to ensure that the observed and simulated modes we are comparing have the same normalisation, we filter the VS simulation with approximate beam transfer functions derived by fitting the window function of a symmetric Gaussian beam to the ratios of $EE$, $BB$ and $TT$ power measured from the Planck data to those measured from the simulations, on scales of $20<l<100$ (below the multipoles where experimental noise becomes significant in the Planck polarisation data). This correction enables a like-for-like comparison on scales of approximately $l<200$, in the case of the GNILC dust data (for which the maps are convolved with a beam with FWHM of $80\,$arcmin to ensure they have the same resolution everywhere), and $l<500$ in the case of the raw 353\,GHz frequency data (approximately the angular scale where polarisation noise in the 353\,GHz channel begins to dominate). The latter is an important range for our purposes, since it encompasses most of the multipole range where Galactic dust dominates over CIB emission in total intensity (see, e.g., \citealt{ref:gnilc}). Furthermore, we verify that, in the 353\,GHz channel, the effects of beam convolution at $l=500$ are still relatively limited and signal power is only suppressed by approximately $10\,\%$.

When using the raw 353\,GHz maps, which include significant amounts of CIB emission, our validation could potentially be confused by its sensitivity to the bispectrum of CIB fields. However, from the calculations in section~\ref{sec:analytic_cib_bias_calc}, we expect this to be small for the expected extent of point-source removal. In order to ensure this is the case, we apply a mask -- also obtained from the \href{http://pla.esac.esa.int/pla/#home}{Planck Legacy Archive} --  which removes point sources detected with $S/N>5$ at 353\,GHz . We do, however, see increased scatter in the bispectrum measurements from the CIB (and the CMB and instumental noise) as discussed below.

\begin{figure}
    \begin{center}
    \includegraphics[width=0.8\columnwidth]{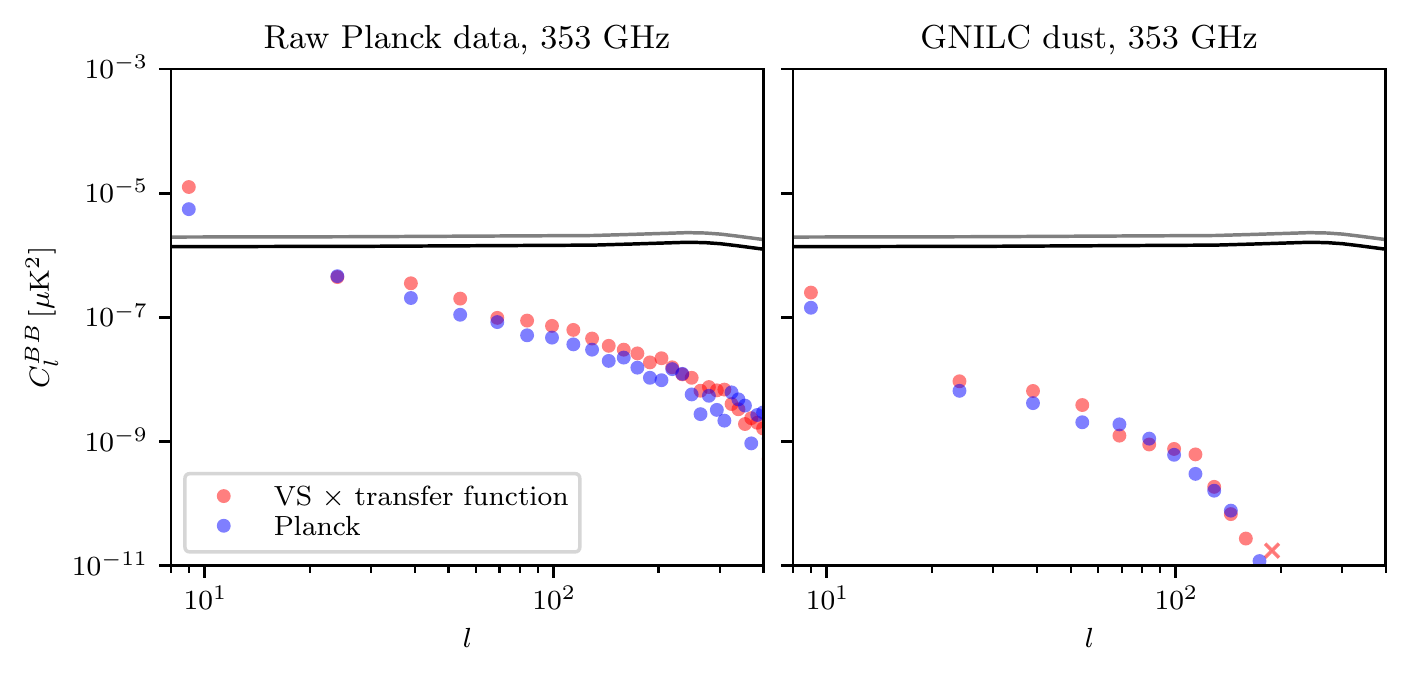}
    \end{center}
    \caption{Delensing bias associated with the bispectrum of Galactic dust arising from a restricted range of multipoles of the VS Galactic dust simulation or the Planck data, after both are filtered and scaled to emulate levels of residual dust expected after foreground cleaning. In order to incorporate the effect of beam convolution, the VS simulation is filtered with a smooth transfer function fit to the power spectra of the data. This enables a comparison in the range $l<200$ when the Planck 353\,GHz GNILC dust data is used to model residuals (right panel) and for $l<500$ when the raw 353\,GHz channel data is used for this purpose instead (left panel). In the latter case, we ensure point-sources are masked. For comparison, the lensing $B$-mode spectrum is plotted before (grey) and after (black) being delensed with the CIB. The amplitude of the dust in each map is scaled down to the level appropriate for the residuals expected in the GNILC estimate of the CIB and the SO LAT and SAT polarisation maps (both after standard ILC-cleaning), as estimated in section~\ref{sec:dust_transfer_funcs}. Furthermore, an $l_{\mathrm{min}}=100$ cut is applied to the fields. Spectra are measured on the ``LB patch'' shown in figure~\ref{fig:planck_60pcsky_mask} and binned with $\Delta l=15$. In both cases, negative values are plotted as circles and positive ones as crosses.}
    \label{fig:compare_bias_vs_vs_planck}
\end{figure}

The results of the delensing bias from the bispectrum of dust computed from the re-scaled VS simulations or the Planck data are shown in figure~\ref{fig:compare_bias_vs_vs_planck}. The bias that appears when using the raw data as input is approximately two orders of magnitude larger than that produced by the GNILC dust maps, owing to the much restricted multipole range going into the latter.

For a given type of input map, the level of agreement between data and simulations is remarkable. The bias calculation based on the raw Planck frequency maps shows higher scatter than seen in the simulation (particularly on smaller scales), which can be explained as the added variance from the other components present in those maps such as the CIB, the CMB or experimental noise. This explanation is consistent with the fact that such increase in scatter is not seen when the GNILC dust products are used. On that note, it is worth pointing out that the similar amplitude and shape of the biases arising from the raw channel data and the pure-dust simulations is consistent with the claim made in section~\ref{sec:analytic_cib_bias_calc} that the $BEI$ bispectrum of the CIB is small once point-sources are masked.

\section{Conclusions}\label{sec:conclusion}
We have investigated ways in which residual foregrounds can impact the power spectrum of CMB $B$-modes after they are delensed using the CIB as a tracer of the matter. We identified two sources of bias which could, in principle, affect constraints on the tensor-to-scalar ratio $r$: residual Galactic dust and residual CIB emission left over in the CMB maps after foreground cleaning.

Given that, at the time of writing, no simulations of polarised CIB emission exist, we calculated the delensing bias arising from higher-point functions of the CIB analytically, in the context of a model where source polarisation angles are uncorrelated. We found that the dominant contribution to this bias ought to be the $\langle BEI\rangle$ bispectrum of the CIB. Our calculation shows that, given the extent of point source masking expected of upcoming experiments, this bias ought to be negligible.
As a side note, our model shows that certain bispectra involving CIB polarisation fields can be non-vanishing even if galaxy polarisation angles are uncorrelated, namely those containing two polarisation legs sourced by the same galaxy. 

We also presented a first attempt to characterise the delensing bias arising from residual Galactic dust left over in CMB and CIB maps used for delensing. We simulated the delensing procedure as applied to these dust residuals,
using a non-Gaussian simulation of Galactic dust from~\citet{ref:vansyngel} filtered to model the expected levels of residuals after foreground cleaning appropriate for CMB $E$- and $B$-mode polarisation and CIB observations in the coming era of CMB experiments. In particular, we estimated residual dust levels for the Simons Observatory's (SO) measurements of $E$-modes with its large-aperture telescope (LAT) and $B$-modes with its small-aperture telescopes, and for the CIB maps of \citet{ref:gnilc} and~\citet{ref:lenz_19}. We also validated our results directly against the Planck data on large and intermediate scales.

We found that the bias from residual Galactic dust is dominated by the $\langle BEI\rangle_{c}$ bispectrum of Galactic dust, and that it is negative. It is also very red, peaking on the very largest angular scales. Given the extent of foreground cleaning expected of upcoming ground-based experiments (we focus on the specifications of SO), the bias only becomes relevant for multipoles $l\lesssim 20 $, scales likely too large to be probed from the ground. Hence, we find that a failure to model this bias at the level of the power spectrum translates to an inference on $r$ that is systematically lower than the truth, but only by an amount smaller than the $1\sigma$ statistical uncertainty. To err on the side of caution, however, we recommend explicitly deprojecting dust from the high-resolution $E$-mode fields -- e.g., with a constrained internal-linear combination approach applied to multi-frequency data~\citet{ref:remazeilles_contrained_ilc} -- used in delensing if the delensed $B$-modes on scales larger than $l<30$ are to be considered. Although this might come at the cost of degraded delensing efficiency, we show that for an experiment with the characteristics of the SO LAT this is a small effect (with only a percent-level increase in the residual lensing power).

On the other hand, the bias could be large enough to be a cause of concern for space-based experiments, such as LiteBird, targeting the reionisation peak of the primary CMB in addition to the recombination peak targeted by ground-based experiments.
By covering larger sky areas, space-based experiments are prone to including regions where the dust is very bright, causing the bias to grow and become significant on scales that the telescope is sensitive to. Furthermore, their situation above the atmosphere allows them to probe the large-scale signal from reionisation, scales where the bias is largest. In spite of this, we show that a combination of dust deprojection in the foreground cleaning of $E$-modes and removal of large-scale CIB modes in the construction of the lensing $B$-mode template is effective in reducing this bias. The latter can always be implemented at little cost in terms of delensing efficiency (at least for $l^{\text{CIB}}_{\text{min}}\lesssim 250$), since internal lensing reconstruction techniques can estimate these large-scale lenses very accurately. Furthermore, the bias is very sensitive to the removal of dust from the CMB polarisation fields involved, so we expect the situation to be less concerning for future satellite missions, for which foreground cleaning will be much improved relative to the levels considered in this paper.
Altogether, we are confident that the bias can be kept in check and CIB delensing can remain a valuable tool.

We note also that, in addition to the biases described above, residual foregrounds ought also to impact the variance of the power spectrum of delensed $B$-modes. However, we defer the study of that problem to future work, given the difficulty of studying it analytically and the lack of realistic simulations to assist a computational treatment.

The results presented in this paper illustrate the community's need for non-Gaussian simulations of Galactic dust extending to small angular scales on which to test the robustness of delensing analyses against the systematic effects identified in this work. Such simulations are also necessary to assess the level of contamination in CMB lensing power spectrum reconstruction from polarisation-based estimators~\citep{ref:CORElensing,ref:beck_2020}. We emphasise that the approach followed to date of testing against Gaussian simulations (see, e.g.,~\citealt{ref:manzotti_delensing} or~\citealt{ref:bicep_delensing}) will not suffice, as validation experiments based on those will be insensitive to the non-Gaussian biases discussed here, which can be significant on the largest angular scales.

\section*{Acknowledgements}
The authors would like to thank Alex Van Engelen, Colin Hill, Byeonghee Yu, Will Coulton and Mathieu Remazeilles for useful discussions and their sharing of products that enabled this work.
ABL gratefully acknowledges support from an Isaac Newton Studentship at the University of Cambridge and the Science and Technology Facilities Council (STFC). AC acknowledges support from the STFC (grant numbers ST/N000927/1 and ST/S000623/1).
TN and BDS acknowledge support from the European Research Council (ERC) under the European Unions Horizon 2020 research and innovation programme (Grant agreement No. 851274). BDS further acknowledges support from an STFC Ernest Rutherford Fellowship. This research used resources of the National Energy Research Scientific Computing Center (NERSC), a U.S. Department of Energy Office of Science User Facility. It also benefited from use of the publicly-available codes \texttt{HealPy}~\citep{ref:healpy_paper}, \texttt{HealPix}~\citep{ref:healpix_paper}, AstroPy~\citep{ref:astropy}, \texttt{CAMB}~\citep{Lewis:1999bs} and \texttt{NaMaster}~\citep{ref:namaster}. This is not an official SO Collaboration paper.

\section*{Data Availability}
The data underlying this article will be shared on reasonable request to the corresponding author. Where the data were provided by a third party (as described in the text), the data will be shared subject to permission from the third party.
\bibliographystyle{mnras}
\bibliography{bibliography}



\appendix

\section{The $BEI$ bispectrum of the CIB}
\label{appendix:BET_bispectrum}
In order to model the $BEI$ bispectrum of the CIB, we begin by considering the specific intensity observed at frequency $\nu$, along a line of sight $\hat{\bmath{n}}$. We shall work in the flat-sky limit, in which case $\hat{\bmath{n}}$ is close to the pole and the 3D comoving position of a source, $\bmath{x}_{\text{3D}}$, can be related to its angular position in the plane of the sky, $\bmath{\theta}$, as $\bmath{x}_{\text{3D}} \approx r(z)(\bmath{\theta}+\hat{\bmath{z}})$, where $r(z)$ is the comoving distance to the redshift where the source is located and $\hat{\bmath{z}}$ is a unit vector along the $z$-direction.
In this limit, the observed temperature is an integral along the line of sight of $j_{\nu}$, the emissivity per unit comoving volume and per solid angle :
\begin{equation}
    I_{\nu}(\bmath{\theta})= \int dz \frac{dr}{dz}a(z)j_{\nu(1+z)}(r(z)(\bmath{\theta}+\hat{\bmath{z}}),z).
\end{equation}
Similarly, the Stokes parameters for linear polarisation can be related to $j^{(\pm)}_{\nu}$, the polarised emissivity per unit comoving volume along the negative $z$-direction, via
\begin{equation}
    (Q\pm iU)_{\nu}(\bmath{\theta})= p \int dz \frac{dr}{dz}a(z)j^{(\pm)}_{\nu(1+z)}(r(z)(\bmath{\theta}+\hat{\bmath{z}}),z),
\end{equation}
where $p$ is the polarisation fraction of the sources, which we will take to be a constant. In the flat-sky limit, the Fourier modes of polarisation can be extracted as
\begin{align}\label{eqn:q_u_multipoles}
    (Q\pm iU)_{\nu}(\bmath{l}) &= \int d^2\bmath{\theta} (Q\pm iU)_{\nu}(\bmath{\theta}) e^{-i\bmath{l}\cdot\bmath{\theta}} \nonumber \\
    & = p \int dz \, \frac{dr}{dz}a(z) \int \frac{d^3\bmath{k}}{(2\pi)^3}\, j^{(\pm)}_{\nu(1+z)}(\bmath{k},z)\int d^2\bmath{\theta} \, e^{ir(z)\bmath{k} \cdot (\bmath{\theta}+\hat{\bmath{z}})} e^{-i\bmath{l}\cdot\bmath{\theta}} \nonumber \\
    & = p \int dz \, \frac{dr}{dz}a(z) \int \frac{d^3\bmath{k}}{2\pi}\, j^{(\pm)}_{\nu(1+z)}(\bmath{k},z)\delta^{(2)}_{\text{D}}(\bmath{k}_\perp r(z) - \bmath{l}) e^{i k_3 r(z)} \nonumber \\
    & = p \int dz \, \frac{dr}{dz}\frac{a(z)}{r^2(z)} \int \frac{d k_3}{2\pi}\, e^{i k_3 r(z)} j^{(\pm)}_{\nu(1+z)}(\bmath{l}/r(z)+k_3 \hat{\bmath{z}},z) ,
\end{align}
and, similarly, for the intensity:
\begin{align}\label{eqn:t_multipoles}
    I_{\nu}(\bmath{l}) & = \int dz \, \frac{dr}{dz}\frac{a(z)}{r^2(z)} \int \frac{d k_3}{(2\pi)} \, e^{i k_3 r(z)} j_{\nu(1+z)}(\bmath{l}/r(z)+k_3 \hat{\bmath{z}},z) .
\end{align}
Here, $\bmath{k}_\perp$ is the projection of the wavevector $\bmath{k}$ perpendicular to the $z$-direction.
We are now in a position to extract $E$ and $B$ modes. In the flat-sky approximation, these can be constructed as
%
\begin{equation}
    (E\pm i B)(\bmath{l}) = -(Q\pm i U)(\bmath{l})e^{\mp 2 i \psi_{\bmath{l}}} \, , \label{eq:EBdef}
\end{equation}
where $\psi_{\bmath{l}}$ is the angle between the Fourier wavevector $\bmath{l}$ and the $x$-direction.
Substituting in from equations~\eqref{eqn:q_u_multipoles} and~\eqref{eqn:t_multipoles}, we obtain
\begin{align}
    E_{\nu}(\bmath{l}) & = -\frac{p}{2}  \int dz \frac{dr}{dz}\frac{a(z)}{r^2(z)} \frac{d k_3}{(2\pi)}e^{i k_3 r(z)} \left[ j^{(+)}_{\nu(1+z)}(\bmath{l}/r(z)+k_3 \hat{\bmath{z}},z) e^{-2i\psi_{\bmath{l}}} + j^{(-)}_{\nu(1+z)}(\bmath{l}/r(z)+k_3 \hat{\bmath{z}},z) e^{2i\psi_{\bmath{l}}}\right]
\end{align}
and
\begin{align}
    B_{\nu}(\bmath{l}) & = \frac{ip}{2} \int dz \frac{dr}{dz}\frac{a(z)}{r^2(z)} \frac{d k_3}{(2\pi)}e^{i k_3 r(z)}  \left[ j^{(+)}_{\nu(1+z)}(\bmath{l}/r(z)+k_3 \hat{\bmath{z}},z) e^{-2i\psi_{\bmath{l}}}  - j^{(-)}_{\nu(1+z)}(\bmath{l}/r(z)+k_3 \hat{\bmath{z}},z) e^{2i\psi_{\bmath{l}}}\right].
\end{align}
The bispectrum can therefore be constructed as 
\begin{align}\label{eqn:long_bispectrum_before_simplifying}
    \langle B_{\nu_1}(\bmath{l}_1) E_{\nu_2}(\bmath{l}_2) I_{\nu_3}(\bmath{l}_3)\rangle  =  \frac{-ip^2}{2 (2\pi)^3} \int & \prod_{j=1}^3 dz_j dk_{j,3}  \frac{dr(z_j)}{dz_j}\frac{a(z_j)}{r^2(z_j)} e^{i k_{j,3} r(z_j)} 
    \bigg[ e^{-2 i (\psi_{\bmath{l}_1}- \psi_{\bmath{l}_2})} \langle j^{(+)}_{\nu_1(1+z_1)}(\bmath{l}_1/r(z_1)+k_{1,3} \hat{\bmath{z}},z_1) \nonumber \\
    & \times j^{(-)}_{\nu_2(1+z_2)}(\bmath{l}_2/r(z_2)+k_{2,3} \hat{\bmath{z}},z_2)
    j_{\nu_3(1+z_3)}(\bmath{l}_3/r(z_3)+k_{3,3} \hat{\bmath{z}},z_3) \rangle \nonumber \\
    & - e^{2 i (\psi_{\bmath{l}_1}- \psi_{\bmath{l}_2})} \langle j^{(-)}_{\nu_1(1+z_1)}(\bmath{l}_1/r(z_1)+k_{1,3} \hat{\bmath{z}},z_1)\nonumber \\
    & \times j^{(+)}_{\nu_2(1+z_2)}(\bmath{l}_2/r(z_2)+k_{2,3} \hat{\bmath{z}},z_2)
    j_{\nu_3(1+z_3)}(\bmath{l}_3/r(z_3)+k_{3,3} \hat{\bmath{z}},z_3) \rangle \bigg],
\end{align}
where we have only retained those terms that survive averaging over independent polarisation directions (see appendix~\ref{sec:equal_z_bispectrum}).

Assuming the 3D bispectrum we have encountered is homogeneous and isotropic, we can write it as
\begin{align}
    \langle \prod_{j=1}^3 j^{(s_j)}_{\nu_j(1+z_j)}(\bmath{l}_j/r(z_j)+k_{j,3} \hat{\bmath{z}},z_j)
    \rangle &=(2\pi)^3\delta^{(3)}_{\text{D}}\left(\bmath{l}_1/r(z_1) + \bmath{l}_2/r(z_2) + \bmath{l}_3/r(z_3) + \left(k_{1,3}+k_{2,3}+k_{3,3}\right) \hat{\bmath{z}}\right) \nonumber\\
    & \quad \times B^{s_1 s_2 s_3}_{\nu_1'\nu_2'\nu_3'}\bigg(\left| \bmath{l}_1/r(z_1) + k_{1,3}\hat{\bmath{z}} \right|,
    \left| \bmath{l}_2/r(z_2) + k_{2,3}\hat{\bmath{z}} \right|,\left| \bmath{l}_3/r(z_3) + k_{3,3}\hat{\bmath{z}} \right|;\, z_1,z_2,z_3 \bigg).
\end{align}
For the sake of compactness, we have introduced the shorthand $s_j$ to label the $(\pm)$ polarisation and the intensity. Additionally, we have defined $\nu' = \nu(1+z)$. The three-dimensional delta function can be factored as
\begin{equation}
\delta^{(3)}_{\text{D}}\left(\bmath{l}_1/r(z_1) + \bmath{l}_2/r(z_2) + \bmath{l}_3/r(z_3) + \left(k_{1,3}+k_{2,3}+k_{3,3}\right)\hat{\bmath{z}}\right)
     = \delta^{(2)}_{\text{D}} \left( \bmath{l}_1/r(z_1) + \bmath{l}_2/r(z_2) + \bmath{l}_3/r(z_3) \right) \delta_{\text{D}} \left(k_{1,3} + k_{2,3} + k_{3,3} \right).
\end{equation}

At this stage, we invoke the Limber approximation (see, e.g.,~\citealt{ref:buchalter_2000} in the context of the projected bispectrum), whereby the radial integrations enforce $k_{j,3}\ll |\bmath{l}_{j}|/r(z_j) $ such that the bispectrum is approximately independent of the $k_{j,3}$. Performing the integrals over the $k_{j,3}$ in equation~\eqref{eqn:long_bispectrum_before_simplifying} gives rise to delta functions in $r_j$, which in turn require that all contributions come from the same redshifts; we find that
\begin{align}\label{eqn:}
    \langle B_{\nu_1}(\bmath{l}_1) E_{\nu_2}(\bmath{l}_2) I_{\nu_3}(\bmath{l}_3)\rangle  = & -\frac{p^2}{2} \sin 2(\psi_{\bmath{l}_1} -\psi_{\bmath{l}_2})(2\pi)^2\delta^{(2)}_{\text{D}}  \left( \bmath{l}_1 + \bmath{l}_2 + \bmath{l}_3 \right) \nonumber \\
    & \times \int dr_1 dr_2 dr_3\, a(z_1)\frac{a(z_2)}{r_2^2} \frac{a(z_3)}{r_3^2} \delta_{\text{D}}\left(r_1 - r_2\right)\delta_{\text{D}}\left(r_1 - r_3\right) B_{\nu_1'\nu_2'\nu_3'}^{(+)(-)()}\left( \frac{l_1}{r_1}, \frac{l_2}{r_2}, \frac{l_3}{r_3};z_1,z_2,z_3 \right) \nonumber \\
    = & -\frac{p^2}{2} \sin 2(\psi_{\bmath{l}_1}-\psi_{\bmath{l}_2}) (2\pi)^2 \delta^{(2)}_{\text{D}}  \left( \bmath{l}_1 + \bmath{l}_2 + \bmath{l}_3 \right) \int dz \frac{dr}{dz}\frac{a^3(z)}{r^4(z)} B_{\nu_1'\nu_2'\nu_3'}^{(+)(-)()}\left( \frac{l_1}{r(z)}, \frac{l_2}{r(z)}, \frac{l_3}{r(z)};z \right)\, ,
\end{align}
where we have used
\begin{equation}
    \delta^{(2)}_{\text{D}}\left(\bmath{l}_1/r(z) + \bmath{l}_2/r(z) + \bmath{l}_3/r(z)\right)
     = r^2(z) \delta^{(2)}_{\text{D}} \left( \bmath{l}_1 + \bmath{l}_2 + \bmath{l}_3 \right) \, .
\end{equation}
The next step is to calculate the equal-redshift 3D galaxy bispectrum. This is done in appendix~\ref{sec:equal_z_bispectrum}. Using its explicit form from equation~\eqref{eqn:equal_z_bispec_ito_S} and using $dr/dz=c/H(z)$ to simplify, we obtain
\begin{align}\label{eqn:cib_bispectrum_general}
    \langle B_{\nu}(\bmath{l}_1) E_{\nu}(\bmath{l}_2) I_{\nu}(\bmath{l}_3)\rangle  = & -\frac{p^2}{2} \sin 2(\psi_{\bmath{l}_1}-\psi_{\bmath{l}_2}) (2\pi)^2 \delta^{(2)}_{\text{D}}  \left( \bmath{l}_1 + \bmath{l}_2 + \bmath{l}_3 \right) \int dz \left[S_\nu^{(3)}(z) + \frac{H(z)}{cr^2(z)}S_\nu^{(2)}(z)S_\nu^{(1)}(z)P_{g}\left(\frac{l_3}{r(z)};z\right)\right],
\end{align}
with $S_\nu^{(n)}(z)$ as defined in equation~\eqref{eqn:Sn_definition}. This expression was derived by assuming that the SED of all sources is the same at a given redshift. By introducing in the integral over redshift appropriate ratios of the SED of the galaxies' luminosity at the redshifted frequencies, it can be generalised to a scenario where the different fields are observed at different frequencies.

\section{Modelling the equal-redshift 3D bispectrum}\label{sec:equal_z_bispectrum}
We now calculate the equal-redshift 3D bispectrum of polarised emissivities under the assumption that all sources have the same SED. For convenience, we consider the case of all three frequencies being the same, which we can easily generalise to mixed frequencies later on by scaling with appropriate ratios across frequency of the sources' SED.

The emissivity at position $\bmath{x}$, redshift $z$ and frequency $\nu$ is
\begin{equation}
j_{\nu}(\bmath{x},z)=\frac{1}{4\pi}\sum_i L_{\nu,i} \delta^{(3)}_{\text{D}}\left(\bmath{x}-\bmath{y}_i\right)
= \frac{1}{4\pi}\sum_i \int d L_\nu \, L_\nu \delta_{\text{D}}\left(L_\nu - L_{\nu,i}\right)\delta^{(3)}_{\text{D}}\left(\bmath{x}-\bmath{y}_i\right)
,
\end{equation}
where $L_{\nu,i}$ is the spectral luminosity of the source labeled by the subscript $i$, and $\bmath{y}_i$ its position, both 
at redshift $z$. Note that the expectation value
\begin{equation}
\langle \delta_{\text{D}}\left(L_\nu - L_{\nu,i}(z)\right)\delta^{(3)}_{\text{D}}\left(\bmath{x}-\bmath{y}_i\right) \rangle = 
n_g(L_\nu,z) ,
\end{equation}
where $n_g(L_\nu,z)\equiv dN/dVdL_\nu$ is the number density of galaxies per comoving volume and specific luminosity range, at redshift $z$. Similarly, the polarised emissivity is
\begin{equation}
j_{\nu}^{(\pm)}(\bmath{x},z)=\frac{1}{4\pi}\sum_i L_{\nu,i} \delta^{(3)}_{\text{D}}\left(\bmath{x}-\bmath{y}_i\right)e^{\pm 2i \theta_i},
\end{equation}
where $\theta$ describes the direction of the linear polarisation in the plane of the sky for emission along the $-z$ direction.
As will immediately become apparent, the only possible independent three-point function involving two factors of polarised emissivity is
\begin{align}\label{eqn:emissivity_bispectrum}
    (4\pi)^3\langle j_{\nu}^{(+)}(\bmath{x}_1,z) j_{\nu}^{(-)}(\bmath{x}_2,z) j_{\nu}(\bmath{x}_3,z) \rangle =& \langle \sum_{ijk} \int dL'_\nu dL''_\nu dL'''_\nu \, L'_{\nu} L''_{\nu} L'''_{\nu} \delta_{\text{D}}\left( L'_\nu-L_{\nu,i} \right)\delta_{\text{D}}\left( L''_\nu-L_{\nu,j} \right)\delta_{\text{D}}\left( L'''_\nu-L_{\nu,k} \right) \nonumber \\
    & \quad \times
    \delta^{(3)}_{\text{D}}\left( \bmath{x}_1 - \bmath{y}_i \right) \delta^{(3)}_{\text{D}}(\bmath{x}_2 - \bmath{y}_j) \delta^{(3)}_{\text{D}}\left( \bmath{x}_3 - \bmath{y}_k \right)  e^{2i(\theta_i-\theta_j)}\rangle \nonumber \\
    & = \delta^{(3)}_{\text{D}}(\bmath{x}_1 - \bmath{x}_2) \int dL'_\nu dL'''_\nu \, \left(L'_\nu\right)^2 L'''_\nu \nonumber \\
    &\quad \times \langle \sum_i \delta^{(3)}_{\text{D}}(\bmath{x}_1 - \bmath{y}_i) \delta_{\text{D}}(L'_\nu-L_{\nu,i})
    \sum_k \delta^{(3)}_{\text{D}}(\bmath{x}_3 - \bmath{y}_k) \delta_{\text{D}}(L'''_\nu-L_{\nu,k}) \rangle .
\end{align}
In going to the last line, we have used the fact that averaging over the polarisation angles $\theta_i$ imposes $\delta_{ij}$. This is to be interpreted as saying that, since the polarisation angle is uncorrelated between sources, the only possible contributions come from terms where a polarisation leg can be matched to another polarisation leg from the same source. This logic has consequences for other interesting bispectra such as $\langle BII\rangle$ or $\langle EII\rangle$, which necessarily vanish. Futhermore, we also learn that any bispectrum with three polarisation legs -- which we might have naïvely expected to accept one-source contributions -- vanish upon averaging as they involve first and third powers of sines and cosines. Of course, this all only holds under the assumption that the polarisation angles of galaxies are uncorrelated.

We can relate the expectation value in equation~\eqref{eqn:emissivity_bispectrum} to the galaxy density and the two-point correlation function; following~\citet{ref:scherrer_and_bertschinger}, we have
\begin{align}
      \langle \sum_i \delta^{(3)}_{\text{D}}(\bmath{x}_1 - \bmath{y}_i) \delta_{\text{D}}(L'_\nu-L_{\nu,i})
    \sum_k \delta^{(3)}_{\text{D}}(\bmath{x}_3 - \bmath{y}_k) \delta_{\text{D}}(L'''_\nu-L_{\nu,k}) \rangle &= n_g(L_\nu',z) \delta^{(3)}_{\text{D}}\left( \bmath{x}_1 - \bmath{x}_3 \right)\delta_{\text{D}}\left( L'_\nu-L'''_\nu \right) \nonumber \\
     & \quad + n_g(L'_\nu,z)n_g(L'''_\nu,z)\left[1+\xi\left(|\bmath{x}_1-\bmath{x}_3|, L'_\nu, L'''_\nu;z\right)\right]\,.
\end{align}
We shall take the galaxy correlation function, $\xi$, to be independent of the luminosity of the galaxies involved from here on.
Using the relation above in equation~\eqref{eqn:emissivity_bispectrum}, we have that
\begin{align}
    \langle j_{\nu}^{(+)}(\bmath{x}_1,z) j_{\nu}^{(-)}(\bmath{x}_2,z) j_{\nu}(\bmath{x}_3,z) \rangle = (4\pi)^{-3} \delta^{(3)}_{\text{D}}\left( \bmath{x}_1 - \bmath{x}_2 \right) \left(\delta^{(3)}_{\text{D}}\left( \bmath{x}_1 - \bmath{x}_3 \right) L_{\nu}^{(3)}(z) + \left[ 1+\xi\left(|\bmath{x}_1-\bmath{x}_3|;z\right)\right]L_{\nu}^{(2)}(z)L_{\nu}^{(1)}(z) \right),
\end{align}
where we have defined
\begin{equation}
  L_{\nu}^{(n)}(z)\equiv \int dL_\nu L_\nu^n \frac{dN}{dVdL_\nu}.
\end{equation}

The bispectrum can be easily obtained from the three-point function by expressing $j_{\nu}(\bmath{x},z)$ in terms of its Fourier transform. This yields
\begin{align}
    (4\pi)^3 \langle j_{\nu}^{(+)}(\bmath{k}_1,z) j_{\nu}^{(-)}(\bmath{k}_2,z) j_{\nu}(\bmath{k}_3,z) \rangle &= \int d^3\bmath{x}_1 d^3\bmath{x}_2 d^3\bmath{x}_3\, \langle j_{\nu}^{(+)}(\bmath{x}_1,z) j_{\nu}^{(-)}(\bmath{x}_2,z) j_{\nu}(\bmath{x}_3,z) \rangle e^{-i\left(\bmath{k}_1\cdot\bmath{x}_1 + \bmath{k}_2\cdot\bmath{x}_2 + \bmath{k}_3\cdot\bmath{x}_3\right)} \nonumber \\
    &= L_{\nu}^{(3)}(z) (2\pi)^3 \delta_{\text{D}}^{(3)}\left(\bmath{k}_1 + \bmath{k}_2 + \bmath{k}_3\right) + L_{\nu}^{(2)}(z)L_{\nu}^{(1)}(z) \biggl[(2\pi)^6 \delta_{\text{D}}^{(3)}\left(\bmath{k}_1 + \bmath{k}_2 \right)\delta_{\text{D}}^{(3)}\left( \bmath{k}_3 \right) \nonumber \\
    & \phantom{\delta_{\text{D}}^{(3)}\left(\bmath{k}_1 + \bmath{k}_2 + \bmath{k}_3\right) + L_{\nu}^{(2)}(z)L_{\nu}^{(1)}(z)} + \int d^3\bmath{x}_1 d^3\bmath{x}_3 \, \xi\left(|\bmath{x}_1-\bmath{x}_3|;z\right)e^{-i\left(\bmath{k}_1 + \bmath{k}_2 \right)\cdot \bmath{x}_1 -i\bmath{k}_3 \cdot \bmath{x}_3} \biggr].
\end{align}
The first term in the last line corresponds to the one-source contribution, and the last term to the two-source. The term involving $\delta^{(3)}_{\text{D}}(\bmath{k}_3)$ arises from a disconnected contraction; we can ignore this term since it will not contribute to the $BEI$ bispectrum for multipoles of $I$ with $l_3 > 0$.
Recalling that the galaxy correlation function, $\xi(\bmath{x})$, is the Fourier transform of the galaxy power spectrum, $P_{g}(k)$, or more explicitly
\begin{equation}
    \xi\left(|\bmath{x}_1-\bmath{x}_3|;z\right) = \int \frac{d^3\bmath{k}}{(2\pi)^3}P_{g}(k;z)e^{i\bmath{k}\cdot \left(\bmath{x}_1 - \bmath{x}_3 \right)},
\end{equation}
we can finally write
\begin{align}\label{eqn:bispectrum_ito_L}
    \langle j_{\nu'}^{(+)}(\bmath{k}_1,z) j_{\nu'}^{(-)}(\bmath{k}_2,z) j_{\nu'}(\bmath{k}_3,z) \rangle = &
    (4\pi)^{-3}(2\pi)^3\delta_{D}^{(3)}\left(\bmath{k}_1 + \bmath{k}_2 + \bmath{k}_3\right) \left[ L_{\nu'}^{(3)}(z) +  L_{\nu'}^{(2)}(z)L_{\nu'}^{(1)}(z)P_{g}(|\bmath{k}_3|; z) \right]\,,
\end{align}
where, recall, $\nu' = \nu(1+z)$
The other non-vanishing term, $\langle j_{\nu'}^{(-)}(\bmath{k}_1,z) j_{\nu'}^{(+)}(\bmath{k}_2,z) j_{\nu'}(\bmath{k}_3,z) \rangle$, has the same value.

It will prove convenient to rewrite this expression in terms of observed flux-densities, $S_\nu$, rather than intrinsic spectral luminosities, $L_{\nu'}$. These are related via the luminosity distance, $d_{L}$, as
\begin{equation}
L_{\nu'} = 4\pi(1+z)^{-1}d_{L}^{2}S_{\nu}.
\end{equation}
In a flat cosmology, the luminosity distance is very simply related to the comoving distance as $d_{L}=(1+z)r$. Hence
\begin{align}
    L_{\nu'}^{(n)}(z) &= \int dL_{\nu'} L^n_{\nu'} \frac{dN}{dVdL_{\nu'}} \nonumber \\
    &= r^{-2}\left(\frac{dr}{dz}\right)^{-1}\int dS_\nu \left(4\pi(1+z)r^2S_{\nu}\right)^n \frac{dN}{dS_\nu dz d\Omega} \nonumber \\
    &= (4\pi)^n (1+z)^n r^{2n-2}\left(\frac{dr}{dz}\right)^{-1} S^{(n)}_{\nu}(z)\,,
\end{align}
with $S^{(n)}_{\nu}(z)$ as defined in equation~\eqref{eqn:Sn_definition} -- a quantity that can be evaluated from observed number counts.

Finally, we can cast equation~\eqref{eqn:bispectrum_ito_L} in the following, more convenient, form
\begin{equation}\label{eqn:equal_z_bispec_ito_S}
    \langle j_{\nu'}^{(+)}(\bmath{k}_1,z) j_{\nu'}^{(-)}(\bmath{k}_2,z) j_{\nu'}(\bmath{k}_3,z) \rangle = 
    (1+z)^3 r^4(z) (2\pi)^3 \delta_{D}^{(3)}\left(\bmath{k}_1 + \bmath{k}_2 + \bmath{k}_3\right) \left(\frac{dr}{dz}\right)^{-1}\left[  S_{\nu}^{(3)}(z) + \frac{1}{r^2(z)}\left(\frac{dr}{dz}\right)^{-1} S_{\nu}^{(2)}(z)S_{\nu}^{(1)}(z)P_{g}(|\bmath{k}_3|; z) \right].
\end{equation}{}

\section{Toy model for CIB higher-point functions}\label{appendix:toy_model_cib_higherpointfns}
In this section, we aim to motivate the claim that, for the extent of point-source removal expected of upcoming CMB experiments, the $EIEI$ trispectrum of the CIB ought to result in a smaller bias than the $BEI$ bispectrum. We do so by studying the one-source contribution in a minimal model where sources locations on the sky are assumed to be random and uncorrelated. Similarly, we assume that source flux-density, polarisation fraction and polarisation angle are all uncorrelated.

If we consider only one-source contributions where all legs in the correlator come from the same galaxy, the biases of equations~\eqref{eqn:cib_bispectrum_bias_firstoccurence} and~\eqref{eqn:cib_trispectrum_bias_general} reduce to
\begin{align}
\label{eqn:BET_bias_1s}
\mathcal{B}^{BEI}_l = 2p^2 G_{353\,\mathrm{GHz}}G^2_{145\,\mathrm{GHz}} \int  \frac{d^2\bmath{l}'}{(2\pi)^2} \frac{\bmath{l}'\cdot (\bmath{l}-\bmath{l}')}{|\bmath{l}-\bmath{l}'|^2} \mathcal{W}^{E}_{l'}\mathcal{W}^{I}_{|\bmath{l}-\bmath{l}'|} \sin^2{2(\psi_{\bmath{l}'}-\psi_{\bmath{l}})} 
\frac{1}{A_{\text{sky}}} \langle \sum_i  S^2_{145\,\text{GHz}, i}S_{353\,\text{GHz}, i} \rangle
\end{align}
and
\begin{multline}
{\mathcal{B}}^{EIEI}_{l,1s} = - 2p^2 G^2_{353\,\mathrm{GHz}}G^2_{145\,\mathrm{GHz}} \frac{1}{A_{\text{sky}}} \langle \sum_i  S^2_{145\,\text{GHz}, i}S^2_{353\,\text{GHz}, i} \rangle \\
\times \int  \frac{d^2\bmath{l}'d^2\bmath{l}''}{(2\pi)^4} \frac{\bmath{l}'\cdot (\bmath{l}-\bmath{l}')}{|\bmath{l}-\bmath{l}'|^2}
\frac{\bmath{l}''\cdot (\bmath{l}+\bmath{l}'')}{|\bmath{l}+\bmath{l}''|^2}\mathcal{W}^{E}_{l'}\mathcal{W}^{E}_{l''}\mathcal{W}^{I}_{|\bmath{l}-\bmath{l}'|}\mathcal{W}^{I}_{|\bmath{l}+\bmath{l}''|} \sin{2(\psi_{\bmath{l}'}-\psi_{\bmath{l}})}\sin{2(\psi_{\bmath{l}''}-\psi_{\bmath{l}})}
\cos2(\psi_{\bmath{l}'}-\psi_{\bmath{l}''})
\, ,
\end{multline}

where the flux density at 145\,GHz, $S_{\text{145\,GHz}}$, is obtained by extrapolating the value at 353\,GHz, $S_{\text{353\,GHz}}$, according to the SED of a thermal dust emitter with temperature $T_{\text{dust}}$ and spectral index $\beta_{\text{dust}}$ as given in table~\ref{tab:constants}. Details of the derivations of the relevant bispectrum and trispectrum are provided in appendices~\ref{appendix:discrete_bispectrum} and~\ref{appendix:ETET_trispectrum}, respectively.

In practice, it is more convenient to compute the sums over sources as integrals over flux density, weighted by the number of sources per steradian per unit flux density, $dN/dS_\nu d\Omega$, in a certain flux-density range $[s_{\nu,\mathrm{min}}, s_{\nu,\mathrm{max}}]$. That is
\begin{equation}
\frac{1}{A_{\text{sky}}} \langle \sum_i  S^n_{\nu, i}\rangle 
 \rightarrow \int_{s_{\nu,\mathrm{min}}}^{s_{\nu,\mathrm{max}}} dS_{\nu} S_{\nu}^{n} \frac{dN}{dS_\nu d\Omega}\, .
\end{equation}

The number density $dN/dS_{\text{353\,GHz}}d\Omega$ can be obtained by fitting a model to sub-mm data. Our parametric forms follow those employed by~\citet{ref:mak_17} and are as follows: in the faint end, a double power-law,
\begin{equation}
    \frac{dN}{dS_{\text{353\,GHz}}d\Omega} = A \left[\left(\frac{S_{\text{353\,GHz}}}{B}\right)^{n_1} + \left(\frac{S_{\text{353\,GHz}}}{B}\right)^{n_2} \right]^{-1}\, ,
\end{equation}{}
with parameters $A=2.82\times10^8\,\mathrm{Jy}^{-1}\,\mathrm{sr}^{-1}$, $B=0.007\,\mathrm{Jy}$, $n_1=6.5$ and $n_2=2.42$; and on the bright end, a single power-law,
\begin{equation}
 \frac{dN}{dS_{\text{353\,GHz}}d\Omega} = k S_{\text{353\,GHz}}^{-2.5}\, ,
\end{equation}{}
with $k=17.24\,\mathrm{Jy}^{1.5}\,\mathrm{sr}^{-1}$.

\begin{figure}
    \begin{center}
    \includegraphics[width=0.6\columnwidth]{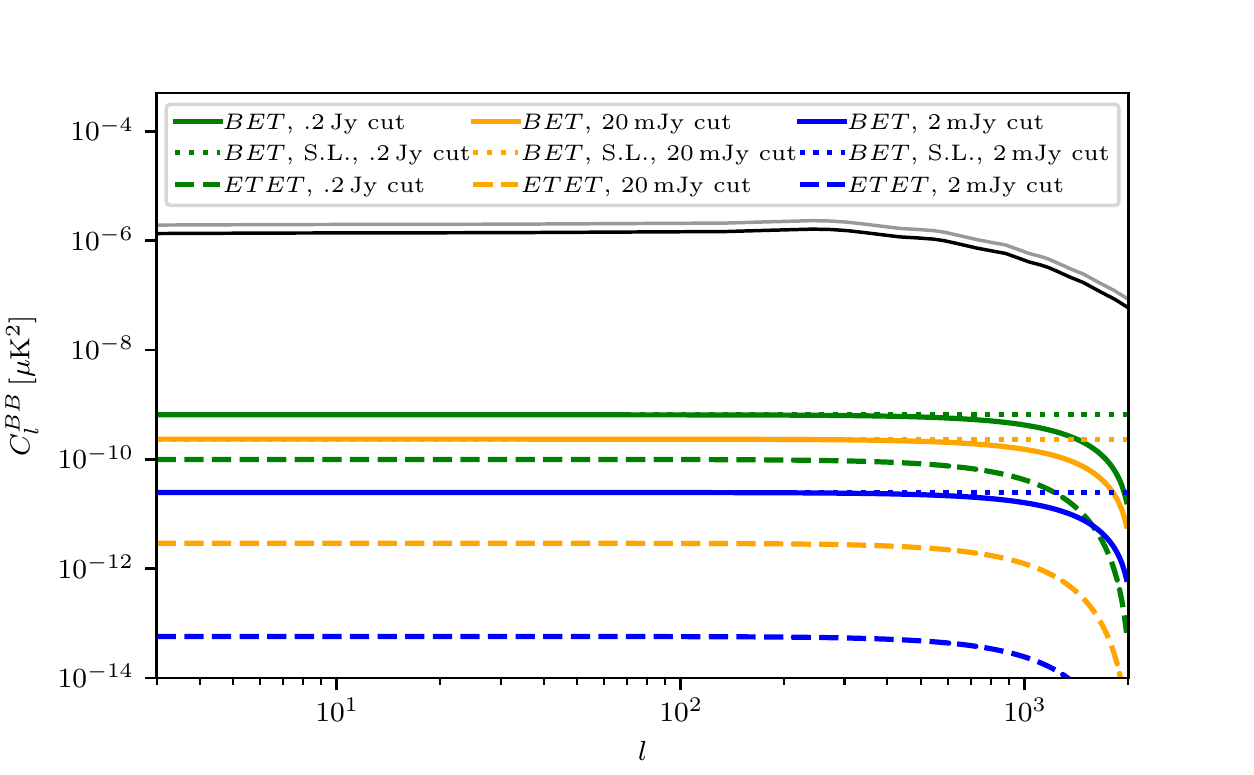}
    \end{center}
    \caption{Delensing biases from the $BEI$ bispectrum (solid) and $EIEI$ trispectrum (dashed) of the CIB, including only 1-source contributions, for point-source masking schemes typical of current and upcoming CMB experiments. The squeezed limit of the bispectra are shown as dotted lines. For comparison, we show also the fiducial power spectra of lensed $B$-modes before (grey) and after delensing with the CIB such that the power is reduced by $30\,\%$ (black).}
    \label{fig:cib_biases_discrete}
\end{figure}

The biases from these one-source terms are plotted in figure~\ref{fig:cib_biases_discrete} as a function of the flux cut. Given that $\mathcal{B}^{EIEI}_{1s}$ scales as the fourth power of the flux density, while $\mathcal{B}^{BEI}_{1s}$ goes only as the third power, the former is bigger than the latter if sources brighter than around $1 \,\mathrm{Jy}$ are left unmasked. However, for the typical flux cut values expected of the next generation of experiments, the bispectrum term dominates over the trispectrum term by several orders of magnitude.

From figure~\ref{fig:cib_biases}, it is clear that, on large scales, the full calculation of $B^{BEI}_{1s}$ is exceptionally well described by a squeezed limit in which a large-scale $B$-mode leg couples to two, small-scale $T$ and a $E$-mode legs. In this limit, equation~\ref{eqn:BET_bias_1s} approaches
\begin{align}
\mathcal{B}^{BEI}_l \approx -p^2 G_{353\,\mathrm{GHz}}G^2_{145\,\mathrm{GHz}} \frac{1}{A_{\text{sky}}} \langle \sum_i  S^2_{145\,\text{GHz}, i}S_{353\,\text{GHz}, i} \rangle \int  \frac{d^2\bmath{l}'}{(2\pi)^2} \mathcal{W}^{E}_{l'}\mathcal{W}^{I}_{l'} \, ,
\end{align}
which is independent of $l$ and thus gives rise to the flat, white-noise-like angular power spectrum we see.

\section{Toy-model CIB bispectrum: one-source term}\label{appendix:discrete_bispectrum}

The one-source contribution to the CIB bispectrum can also be calculated simply by working directly with the flux densities of the sources. Consider a galaxy at angular position $\bmath{\theta}_i$, with flux-density $S_{\nu,i}$ at observed frequency $\nu$. This contributes to the specific intensity of the CIB as
\begin{equation}
    I_\nu(\bmath{\theta}) = S_{\nu,i} \delta_{\mathrm{D}}^{(2)}(\bmath{\theta}-\bmath{\theta}_i)\, .
\end{equation}
In the Fourier domain, this is
\begin{equation}
    I_\nu(\bmath{l}) = S_{\nu,i} e^{-i\bmath{l}\cdot \bmath{\theta}_i}\,.
\end{equation}
For $N$ such sources this generalises to
\begin{equation}
    \label{eqn:temperature_FT_N_sources}
    I_\nu(\bmath{l}) = \sum_{i=1}^N S_{\nu,i} e^{-i \bmath{l}\cdot \bmath{\theta}_i} \,.
\end{equation}
If we allow the sources to be linearly polarised with a polarisation fraction $p$ -- which we shall assume is independent of flux and position -- the observed polarisation can be described by the Stokes parameters
\begin{equation}
    (Q\pm i U)_\nu(\bmath{l}) = \sum_i p S_{\nu,i} e^{\pm 2i\chi_i}  e^{-i \bmath{l}\cdot \bmath{\theta}_i} \, ,
\end{equation}
where $\chi_i$ is the angle between the polarisation direction of the source and the $x$-axis. (We use $\chi$ here rather than $\theta$ to avoid confusion with the angular position of the source.)
The $E$- and $B$-modes of the observed polarisation follow from equation~\eqref{eq:EBdef}:
\begin{align}
    E_\nu(\bmath{l}) &= - p \sum_i S_{\nu,i} e^{-i\bmath{l}\cdot \bmath{\theta}_i} \cos 2 (\psi_{\bmath{l}} - \chi_i) \, , \\
    B_\nu(\bmath{l}) &= p \sum_i S_{\nu,i} e^{-i\bmath{l}\cdot \bmath{\theta}_i} \sin 2 (\psi_{\bmath{l}} - \chi_i) \, .
\end{align}
The bispectrum at frequency $\nu$ can therefore be constructed as
\begin{align}
    \langle B_\nu(\bmath{l}_1) E_\nu(\bmath{l}_2) I_\nu(\bmath{l}_3)\rangle  &= - p^2 \langle \sum_{ijk} S_{\nu,i} S_{\nu,j} S_{\nu,k} e^{-i\bmath{l}_1\cdot \bmath{\theta}_i} e^{-i\bmath{l}_2\cdot \bmath{\theta}_j} e^{-i\bmath{l}_3\cdot \bmath{\theta}_k} 
    \sin 2 (\psi_{\bmath{l}_1} - \chi_i) \cos 2 (\psi_{\bmath{l}_2} - \chi_j) \rangle \, .
\end{align}

If we assume that the polarisation angles $\chi_i$ are independent of the positions and fluxes of sources, averaging over the angles will cause the bispectrum to vanish unless $i=j$. This implies that either $i=j=k$ or $i=j\neq k$. If we ignore source clustering, so that the positions of the sources are independent of each other, the case $i=j\neq k$ gives vanishing contribution for $\bmath{l}_3 \neq 0$. Keeping only the one-source term, $i=j=k$, we have
\begin{align}
    \langle B_\nu(\bmath{l}_1) E_\nu(\bmath{l}_2) I_\nu(\bmath{l}_3)\rangle  &= - p^2 \langle \sum_{i} S_{\nu,i}^3 e^{-i(\bmath{l}_1+\bmath{l}_2+\bmath{l}_3)\cdot \bmath{\theta}_i}
    \sin 2 (\psi_{\bmath{l}_1} - \chi_i) \cos 2 (\psi_{\bmath{l}_2} - \chi_i) \rangle \nonumber \\
    &= - \frac{p^2}{2} \sin 2 (\psi_{\bmath{l}_1}-\psi_{\bmath{l}_2})\langle \sum_{i} S_{\nu,i}^3 e^{-i(\bmath{l}_1+\bmath{l}_2+\bmath{l}_3)\cdot \bmath{\theta}_i} \rangle \, ,
\end{align}
where we have averaged over the polarisation angles in passing to the second line. Finally, averaging over the source locations across a sky area $A_{\text{sky}}$ gives
\begin{align}\label{eqn:general_bispectrum_final}
    \langle B_\nu(\bmath{l}_1) E_\nu(\bmath{l}_2) I_\nu(\bmath{l}_3)\rangle  &= - \frac{p^2}{2} \sin 2 (\psi_{\bmath{l}_1}-\psi_{\bmath{l}_2}) (2\pi)^2 \delta_{\text{D}}^{(2)}(\bmath{l}_1+\bmath{l}_2+\bmath{l}_3)
     \frac{1}{A_{\text{sky}}} \langle \sum_i S_{\nu,i}^3 \rangle \, .
\end{align}
It can be shown that this spectrum satisfies all the required symmetries. Note also that it agrees with the one-source component of equation~\eqref{eqn:cib_bispectrum_general} on replacing
\begin{equation}
\frac{1}{A_{\text{sky}}} \langle \sum_i S_{\nu,i}^3 \rangle \rightarrow \int dz\, \int dS_\nu \, S_\nu^3 \frac{dN}{d S_\nu dz d\Omega}
= \int dz\, S^{(3)}_\nu(z) \, .
\end{equation}

\section{Toy-model CIB trispectrum: one-source term}
\label{appendix:ETET_trispectrum}
Following the calculation of appendix~\ref{appendix:discrete_bispectrum}, we calculate here the $EIEI$ trispectrum of the CIB. Assuming once again that flux-density, position and polarisation are uncorrelated, the full four-point function is given by
\begin{align}\label{eqn:general_trispectrum}
    \langle E_\nu(\bmath{l}_1) E_\nu(\bmath{l}_2) I_\nu(\bmath{l}_3)I_\nu(\bmath{l}_4)\rangle = & \frac{1}{2}p^2
    \cos{2(\psi_{\bmath{l}_1}-\psi_{\bmath{l}_2})} \langle \sum_{ikm} S^2_{\nu,i} S_{\nu,k} S_{\nu,m} e^{-i (\bmath{l}_1 + \bmath{l}_2)\cdot \bmath{\theta}_i}
    e^{-i \bmath{l}_3 \cdot \bmath{\theta}_k} e^{-i\bmath{l}_4\cdot \bmath{\theta}_m}\rangle \,.
\end{align}
Once again, we have the restriction that both polarisation legs must come from the same source, though this time there can in principle also be two-source and three-source contributions. The one-source contribution to the \emph{connected} four-point function (trispectrum) is
\begin{align}
    \langle E_\nu(\bmath{l}_1) E_\nu(\bmath{l}_2) I_\nu(\bmath{l}_3)I_\nu(\bmath{l}_4)\rangle_{c}  &= \frac{1}{2} p^2
    \cos{2(\psi_{\bmath{l}_1}-\psi_{\bmath{l}_2})} (2\pi)^2\delta^{(2)}_{\mathrm{D}}(\bmath{l}_1 + \bmath{l}_2 + \bmath{l}_3 + \bmath{l}_4)
    \frac{1}{A_{\text{sky}}}\langle \sum_i S^4_{\nu,i} \rangle\,.
\end{align}

\section{Simulating Gaussian LSS tracers correlated with CMB lensing}\label{appendix:simulating_correlated_tracers}
     Though in this paper we are only concerned with the CIB, we present here a more general framework for generating simulations of large-scale-structure tracers which are appropriately correlated with a given simulation of the CMB lensing potential and amongst themselves. The basic idea --- which has been studied extensively in the literature around Cholesky decompositions --- is to set up a linear system of equations from which it will be possible to solve for the coefficients involved granted we know the theoretical auto- and cross- spectra of the tracers.\par
 
     Suppose we are to work with just three tracers: the CIB, LSST galaxies and an internal reconstruction, with spherical harmonic coefficients expressed as $I_{lm},g_{lm},\kappa^{rec}_{lm}$, respectively. Given a map of the true convergence with spherical harmonic coefficients $\kappa_{lm}$, the different tracers are correlated as 
     \begin{align}
         &\kappa^{rec}_{lm} = \kappa_{lm} + n_{lm}\,,\\
         &g_{lm} = A_l^{g\kappa}\kappa_{lm} + u_{lm}\,,\\
         &I_{lm} = A_l^{I\kappa}\kappa_{lm} + A_l^{gI}u_{lm} + e_{lm}\,,\\
     \end{align}
     where $ n_{lm},u_{lm},e_{lm}$ are coefficients of the noise -- and as such are presumed to each be uncorrelated with everything else -- and $ \kappa_{lm} $ are the coefficients of the true convergence for the particular realisation that we wish to generate correlated tracers for. We can solve for the coefficients
     \begin{align}
         &A_l^{g\kappa} = \frac{C_l^{g\kappa}}{C_l^{\kappa\kappa}}\,,\\
         &A_l^{I\kappa} = \frac{C_l^{I\kappa}}{C_l^{\kappa\kappa}}\,,\\
         &A_l^{gI} = \frac{C_l^{gI}-A_l^{g\kappa}A_l^{I\kappa}C_l^{\kappa\kappa}}{C_l^{uu}}
     \end{align}
     and the noise spectra
     \begin{align}
         &C_l^{nn} = N_l^{\kappa\kappa}\,,\\
         &C_l^{uu} = C_l^{gg} - (A_l^{g\kappa} )^2C_l^{\kappa\kappa}\,,\\
         &C_l^{ee} = C_l^{II} - (A_l^{I\kappa} )^2C_l^{\kappa\kappa}-(A_l^{gI} )^2C_l^{uu}\,,
     \end{align}
     where $N_l^{\kappa\kappa}$ is the internal reconstruction noise level.\par

    From the above, it is clear that the problem is soluble if we know all the auto- and cross- spectra, in which case all we have to do is draw the coefficients $ n_{lm},u_{lm}$ and $e_{lm}$ from probability distributions with angular power spectra $ N_{l}^{\kappa\kappa},C_{l}^{uu}$ and $C_l^{ee}$.\par
    
    The solution above can be generalised to any number of tracers as follows. Consider tracer $i$ with auto- spectrum $C^{ii}$ and cross-spectrum $C^{ij} $ with tracer $j$. Let it be described as a linear combination $\sum_p \sum_{lm} A_l^{ij} a^p(l,m)$ of harmonic coefficients $a^p$ with angular power spectra $C_l^{a^{p}a^{p}}$ and scaled with weights given by the square, off-diagonal matrix $A_l^{ij}$. In the case above, $a^p=\{\kappa_{lm}, u_{lm}, e_{lm}\}$, $C_l^{a^{p}a^{p}} = \{C_l^{\kappa \kappa}, C_l^{uu}, C_l^{ee} \}$. The general formula for the weights is
    \begin{equation}
        A_l^{ij} = \frac{1}{C_l^{a^{j}a^{j}}}\bigg( C^{ij}_l-  \sum_{p=0}^{j-1}  A_l^{jp} A_l^{ip}C_l^{a^{p}a^{p}}\bigg)\,,
    \end{equation}
    with auxiliary spectra given by:
    \begin{equation}
    C_l^{a^{j}a^{j}} = C^{jj}_l - \sum_{p=0}^{j-1} \big(A_l^{jp}\big)^2 C_l^{a^{p}a^{p}}\,.
    \end{equation}
This infrastructure is implemented and made publicly-available on \texttt{GitHub}\footnote{\texttt{https://github.com/abaleato/MultitracerSims4Delensing}}.

\section{Constructing a curved-sky B-mode template}\label{sec:curved_sky_template}
We describe here the construction of a map-level template for lensing B-modes. We work to leading order in lensing and on the curved-sky formalism, a combination that was shown by~\citet{ref:challinor_05} to be a very good approximation to the true B-modes on large angular scales. We arrive at a fast position-space implementation which is made publicly-available on \texttt{GitHub}\footnote{\texttt{https://github.com/abaleato/curved\_sky\_B\_template}}.  These lensed B-modes can be approximated in harmonic space by\footnote{In this notation summation is implicit over matching pairs of indices.}
\begin{align}
     \tilde{B}_{lm} &= \sum_{(lm)_1} \sum_{(lm)_2}  \phi_{(lm)_1} E_{(lm)_2} \frac{1}{2i} \big[ _{2}I_{l l_1 l_2}^{m m_1 m_2} - _{-2}I_{l l_1 l_2}^{m m_1 m_2}\big] \nonumber\\
     &=  \sum_{(lm)_1} \sum_{(lm)_2} \frac{-i}{2} \frac{1}{2}(-1)^m \phi_{(lm)_1} E_{(lm)_2}  \sqrt{(2l+1)(2l_1+1)(2l_2+1)/{4\pi}} [l_1(l_1+1)+l_2(l_2+1)-l(l+1)]\nonumber\\
     & \quad \times \begin{pmatrix}l_1&l_2&l \\ m_1&m_2&-m \end{pmatrix} \bigg[ \begin{pmatrix}l_1&l_2&l \\ 0&-2&2 \end{pmatrix} -  \begin{pmatrix}l_1&l_2&l \\ 0&2&-2 \end{pmatrix} \bigg]\,.
\end{align}
Inspired by the implementation of the quadratic estimators for lensing reconstruction in the publicly-available code \texttt{QuickLens}\footnote{\texttt{https://github.com/dhanson/quicklens}, though an amended and extended version can be found in the alternative repository \texttt{https://github.com/abaleato/Quicklens-with-fixes}}, we write
\begin{align}
    \hat{\tilde{B}}_{lm} = \frac{(-1)^{m}}{2} \sum_{(lm)_1}\sum_{(lm)_2} \begin{pmatrix}l_1&l_2&l \\ m_1&m_2&-m \end{pmatrix} W_{l_1 l_2 l} \hat{\phi}_{(lm)_1} \hat{E}_{(lm)_2}\,,
\end{align}
where the weights
\begin{align}
    W_{l_1 l_2 l} & = \frac{-i}{2} \sqrt{(2l+1)(2l_1+1)(2l_2+1)/{4\pi}}  [l_1(l_1+1)+l_2(l_2+1)-l(l+1)] \bigg[ \begin{pmatrix}l_1&l_2&l \\ 0&-2&2 \end{pmatrix} -  \begin{pmatrix}l_1&l_2&l \\ 0&2&-2 \end{pmatrix} \bigg]
\end{align}
can be cast in separable form as
\begin{align}
    W_{l_1 l_2 l} = \sum_i W^{i}_{l_1 l_2 l}\,,
\end{align}
with 
\begin{align}
    W^{i}_{l_1 l_2 l} = \sqrt{(2l+1)(2l_1+1)(2l_2+1)/{4\pi}} \begin{pmatrix}l_1&l_2&l \\ -s_1^i&-s^i_2&s \end{pmatrix} w^i_{l_1} w^i_{l_2} w^i_{l}\,.
\end{align}
The value for the separable weights $w^i_{l_j}$ can be found in Table~\ref{tab:weights}.
\begin{table}
    \centering
    \begin{tabular}{| c | c  c  c  c  c  c |}
        \hline
        $i$& $s_1^i$& $s_2^i$& $s^i$& $w_{l_1}^i$ &$w_{l_2}^i$ & $w_{l}^i$\\ \hline
        $1$ & $0$ &$2$ &$2$& $l_1(l_1+1)$ & $-1/2$ & $i$ \\ 
        $2$ & $0$ &$2$ &$2$& $-1/2$ & $l_2(l_2+1)$ & $i$ \\ 
        $3$ & $0$ &$2$ &$2$& $1/2$ & $i$ & $l(l+1)$ \\ 
        $4$ & $0$ &$-2$ &$-2$& $l_1(l_1+1)$ & $1/2$ & $i$ \\ 
        $5$ & $0$ &$-2$ &$-2$& $1/2$ & $l_2(l_2+1)$ & $i$ \\ 
        $6$ & $0$ &$-2$ &$-2$& $-1/2$ & $i$ & $l(l+1)$ \\
        \hline
    \end{tabular}
    \caption{Weights for a fast separable implementation of the lensed $B$-mode template.}\label{tab:weights}
\end{table}

\bsp	
\label{lastpage}
\end{document}